\documentclass[11pt]{article}
	
	\newcommand{\blind}{0}

    \makeatletter
    \renewcommand\section{\@startsection {section}{1}{\z@}%
                                       {-3.5ex \@plus -1ex \@minus -.2ex}%
                                       {2.3ex \@plus.2ex}%
                                       {\normalfont\fontfamily{phv}\fontsize{16}{19}\bfseries}}
    \renewcommand\subsection{\@startsection{subsection}{2}{\z@}%
                                         {-3.25ex\@plus -1ex \@minus -.2ex}%
                                         {1.5ex \@plus .2ex}%
                                         {\normalfont\fontfamily{phv}\fontsize{14}{17}\bfseries}}
    \renewcommand\subsubsection{\@startsection{subsubsection}{3}{\z@}%
                                        {-3.25ex\@plus -1ex \@minus -.2ex}%
                                         {1.5ex \@plus .2ex}%
                                         {\normalfont\normalsize\fontfamily{phv}\fontsize{14}{17}\selectfont}}
    \makeatother
	\usepackage{amsmath}
	\usepackage{graphicx}
	\usepackage{enumerate}
	\usepackage[natbibapa]{apacite} % APA 7th style (checklist item 13); keeps \citet/\citep
	\usepackage{url} % not crucial - just used below for the URL
    \usepackage{tabularx}
    \usepackage{longtable}
    \usepackage{placeins}
    \usepackage{amssymb}\usepackage{mathtools}\allowdisplaybreaks
	\usepackage{array}
    \usepackage{booktabs}
    \usepackage[table,xcdraw]{xcolor}
    \usepackage{amsthm}
    \usepackage{wrapfig}
    \usepackage[font=normalsize,labelfont=bf]{caption} % captions at 11 pt (checklist item 7)

\begin{document}
		
			%%%%%%%%%%%%%%%%%%%%%%%%%%%%%%%%%%%%%%%%%%%%%%%%%%%%%%%%%%%%%%%%%%%%%%%%%%%%%%
		\def\spacingset#1{\renewcommand{\baselinestretch}%
			{#1}\small\normalsize} \spacingset{1}
		%%%%%%%%%%%%%%%%%%%%%%%%%%%%%%%%%%%%%%%%%%%%%%%%%%%%%%%%%%%%%%%%%%%%%%%%%%%%%%
		
		\if0\blind
		{
			% \title{\bf {A Stochastic Optimization Model for Scheduling Tenders for Pediatric Vaccines in Low- and Low-Middle-Income Countries}}
   			\title{\bf {Long Term Pediatric Vaccine Tender Scheduling with Quantity Discounts}}
			\author{Nicholas Uhorchak$^a$, Ruben A. Proano$^b$, Sandra Eksioglu$^a$, Burak Eksioglu$^c$\\
			$^a$ Department of Industrial Engineering, University of Arkansas \\
            $^b$ Department of Industrial and Systems Engineering,\\
                    Rochester Institute of Technology \\
             $^c$ Department of Industrial, Manufacturing and Systems Engineering,\\
                    Texas Tech University}
			\date{}
			\maketitle
		} \fi
		
		\if1\blind
		{

            \title{\bf Long Term Pediatric Vaccine Tender Scheduling with Quantity Discounts}
			\author{Author information is purposely removed for double-blind review}
			
\bigskip
			\bigskip
			\bigskip
			\begin{center}
				{\LARGE\bf Long Term Pediatric Vaccine Tender Scheduling with Quantity Discounts}
			\end{center}
			\medskip
		} \fi
		\bigskip

\begin{abstract}
Pediatric vaccine procurement for low- and lower-middle-income countries (LICs and LMICs) requires long-term coordination across heterogeneous vaccine markets, competing stakeholder objectives, and uncertain demand, yet tenders are currently scheduled as a series of independent renewals. We extend the Hypothetically Coordinated Vaccine Market (HCVM) framework to schedule multi-year tenders with flexible lengths and all-unit quantity discounts, and compare three objectives: a balanced coordinating-entity model (UG), a social-benefit model that minimizes missed doses (SB), and a producer profit-maximizing model (MP). Using UNICEF ten-year demand forecasts and 30 demand replications across six discount scenarios, we find that the procurement objective, not demand variability, drives cost-to-tender and schedule complexity. Quantity discounts favor large producers, concentrating the market and increasing supply chain vulnerability. Discount effectiveness depends on market maturity rather than valency: pentavalent and hexavalent vaccines are already near their price floor, and among monovalent vaccines only rotavirus responds to discounts, whereas HPV and PCV remain too expensive to compete. Removing pentavalent and hexavalent vaccines from the procurement pool sharply increases missed doses. These results give planners practical guidance on objective selection, discount design, and protecting access to combination vaccines.
\end{abstract}
			
	\noindent%

	%\newpage
	\spacingset{1.5} % DON'T change the spacing!

\section{Introduction} \label{s:intro}

Pediatric immunization saves millions of children's lives each year \citep{WHO2024}, yet persistent coverage gaps remain in low- and lower-middle-income countries (LICs and LMICs), where financial constraints, supply limitations, and fragmented procurement hinder access to affordable vaccines \citep{Dimitrova2023, Wang2025}. Whether vaccines reach these children depends jointly on long-term planning, supply coordination, market dynamics, and pricing policy.

To address these challenges, coordinating entities such as the United Nations Children's Fund (UNICEF), Gavi, the Vaccine Alliance (Gavi), and the Pan American Health Organization (PAHO) pool demand, negotiate prices, and facilitate access on behalf of eligible countries. They operate in a global market supplied by a few large multinational producers and a growing number of developing-country producers. In LIC/LMIC markets, UNICEF acts as the central procurement agent through competitive tenders, and the resulting economies of scale yield the lowest prices per dose; eligible countries may also receive Gavi co-financing.

\begin{wrapfigure}{l}{0.35\textwidth}
    \centering
    \includegraphics[width=0.35\textwidth]{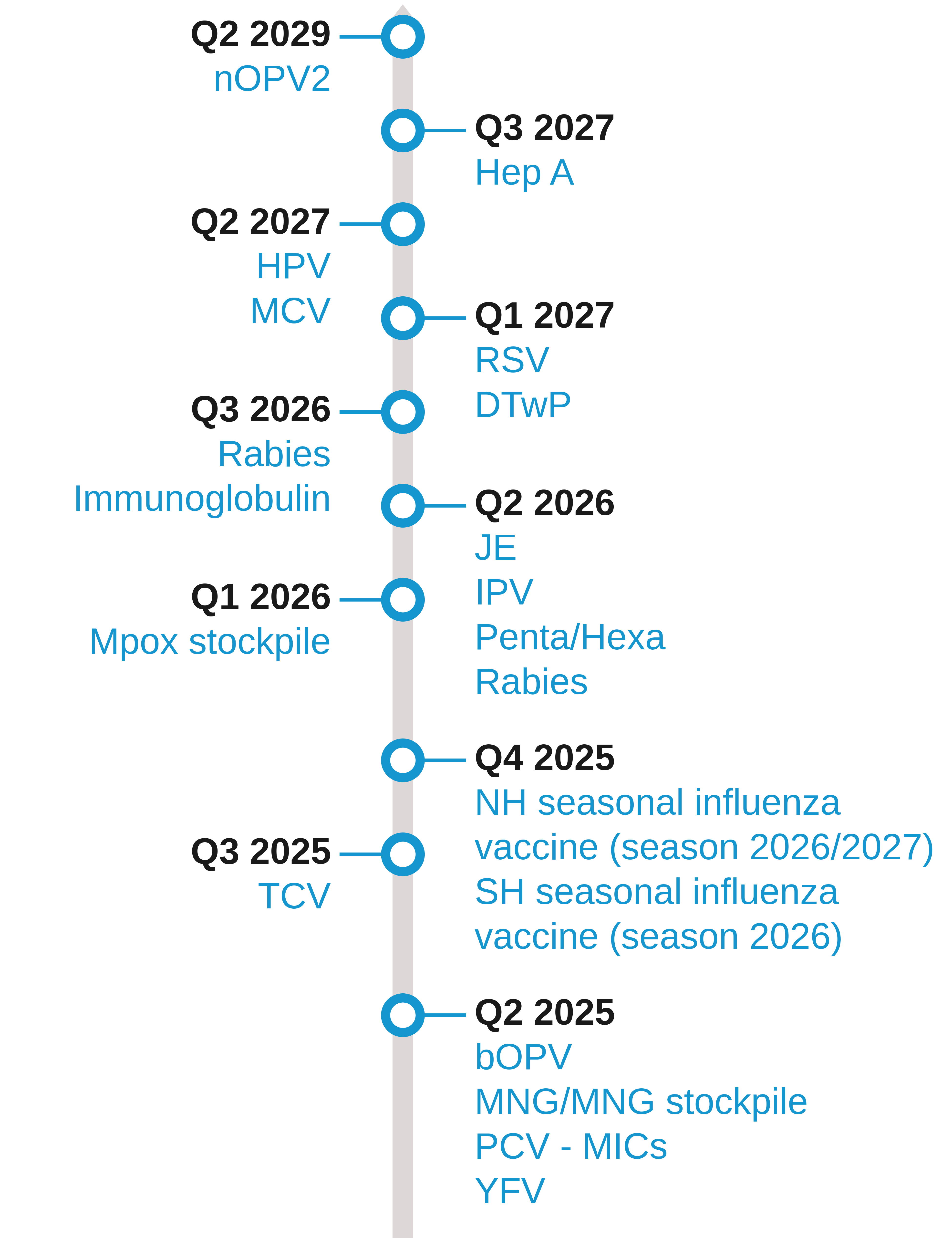}
    \caption{UNICEF Vaccine Tender Schedule: 2024–2027. Source: \citet{UNICEF_Vax_Tender}.}
    \label{fig:UNICEF_vax_tender_sched}
\end{wrapfigure}

A vaccine tender is a contractual process in which producers commit to supply specific quantities at an agreed price over a predetermined duration (the tender length). Procurement follows rolling, renewable frameworks such as the UNICEF tender schedule in Figure~\ref{fig:UNICEF_vax_tender_sched}. Gavi structures market shaping around multi-year strategic periods but continuously revises supply arrangements and demand forecasts \citep{gavi2018vis, gavi2023market}, and PAHO's Revolving Fund relies on periodic tenders and annually updated country demand \citep{paho2020rf}. Although tenders may last up to five years, their scheduling is not governed by a long-term planning horizon: each tender is re-tendered at or near expiration rather than planned as part of a coordinated multi-year sequence.

The Hypothetically Coordinated Vaccine Market (HCVM) framework analyzes how coordinating entities negotiate quantities and prices across heterogeneous country segments to improve affordability for buyers and profitability for producers \citep{Proano2012, AlvesMaciel2022, alves2024enhancing}. These studies, however, do not consider multi-year tenders or their long-term scheduling, and the broader literature on vaccine tender scheduling is sparse. This gap motivates the following research questions:

\noindent {\bf RQ1:}  How does multi-year tender scheduling with flexible tender lengths affect missed doses, cost, and schedule complexity, and how do these outcomes vary with the procurement objective and vaccine type?

\noindent {\bf RQ2:} How do quantity-based discount structures influence producer participation, and what are their consequences for costs and missed doses?

\noindent {\bf RQ3:} How sensitive is coverage to a sustained loss of combination vaccine supply, and can lower-valency alternatives compensate?

We answer these questions by extending the HCVM to schedule multi-year tenders of one to five years with all-unit quantity discounts, under three procurement objectives. We also use the model as a sensitivity analysis tool, evaluating a hypothetical sustained loss of pentavalent and hexavalent supply and the discounts needed to draw three monovalent vaccines into the market.

The remainder of this paper is structured as follows. Section~\ref{s:LitReview} reviews the relevant literature, Section~\ref{s:Models} presents the models and data, Section~\ref{s:Analysis} analyzes the results, and Section~\ref{s:Conclusion} concludes.

\section{Literature Review and Contributions} \label{s:LitReview}

The vaccine literature spans development \citep{Fialho2023}, public health impact \citep{Hyde2012}, distribution logistics \citep{Duijzer2018}, procurement policy \citep{Gianfredi2021}, and market coordination \citep{Gui2026}; yet the operational and market mechanisms that jointly govern tender scheduling, procurement, and pricing remain fragmented and underexplored. Vaccine tenders are the mechanism through which coordinating entities and producers enter long-term procurement agreements to meet projected demand \citep{UNICEF_supply}. Work on tender scheduling and on immunization program planning, including routine Expanded Programme on Immunization (EPI) activities and Supplementary Immunization Activities (SIAs), rarely intersects; \citet{uhorchak_vax_taxonomy} provide a taxonomic review of this intersection.

Tendering and competitive bidding have a long theoretical history, from the foundational work on auction design, bidder strategy, and revenue equivalence \citep{Vickrey1961, Holt1980, Milgrom, Klemperer} to studies of practical tender design, including the trade-off between bidding and negotiation and the role of contract duration \citep{Bajari2003, MacKay2022}. Applied studies examine tenders in road maintenance \citep{Yarmukhamedov2020} and public construction \citep{Hanak2021}. These works rely on empirical, econometric, or game-theoretic analysis and do not address long-term tender scheduling.

Tender scheduling is most closely related to multi-period procurement and lot-sizing problems, in which order timing and quantities are chosen over a finite horizon to satisfy demand at minimum cost \citep{Wagner1958, Brahimi2017}. In the vaccine context, a tender fixes the price and quantity at which producers supply purchasing countries, which then place procurement orders (shipments from a producer to a country) at the agreed terms. The scheduling decision determines when tenders start, how long they last, and how procurement is structured over time, mirroring the period, lot-size, and coverage decisions of lot-sizing models. It is distinct from immunization scheduling, which concerns the timing of vaccine administration to individual patients \citep{Engineer2009, Smalley2011}. Dynamic programming solves single-item, single-supplier lot-sizing problems optimally \citep{Wagner1958}, but the multi-item capacitated problem is NP-hard \citep{Bitran1982}, and multi-antigen tender scheduling inherits this complexity. Exact methods such as branch-and-bound suit small instances \citep{Brahimi2017}, while heuristics such as Lagrangian relaxation, simulated annealing, and genetic algorithms are used for larger ones \citep{Holland1975, Held1971, Kirkpatrick1983}.

Quantity discounts have been studied extensively under both all-unit and incremental structures \citep{Benton1996, Pereira2015}, and they act not only as cost-reduction mechanisms but as strategic instruments that shape vendor participation and buyer--seller coordination \citep{Monahan1984, MUNSON1998}. Embedding them in a multi-antigen tender scheduling mixed-integer program (MIP) adds binary variables and cost breakpoints that compound the problem's combinatorial complexity.

Within operations research, vaccine market mechanisms have been studied mainly through the HCVM framework. \citet{Proano2012} propose a nonlinear MIP for pricing antigen bundles; \citet{AlvesMaciel2022} optimize market-segment-to-entity assignments to balance affordability and producer profitability; and \citet{alves2024enhancing} use sequential bargaining to examine how coordination structure, tiered segmentation, fixed-cost recovery, and procurement ordering affect these outcomes, treating tenders as single-period procurement events. \citet{Meznarich2026} examine pricing and quantities of Diphtheria-Tetanus-Pertussis-family vaccines within a single-year tender through a game-theoretic lens. None of these studies schedules tenders over a multi-year horizon.

\paragraph{Key contribution of this study}

Current practice publishes tender calendars years in advance, but each tender is scheduled around the expiration of its own contract, so the calendar is a series of independent renewals rather than a holistic plan across products and periods. We address this gap within the HCVM framework and make the following contributions.

First, we extend the operational planning horizon for vaccine tender scheduling in the HCVM's LIC/LMIC segments, with flexible tender lengths of one to five years. We show that schedule complexity varies with the procurement objective and vaccine type, with multivalent vaccines scheduled differently from univalent ones.

Second, we introduce quantity-based discount structures into the HCVM and examine how buyer- and producer-side discounts shift coordinating-entity costs, production costs, and missed doses. Poorly designed discounts can misalign stakeholder incentives and undermine vaccine availability.

Third, we show that quantity discounts favor large producers, concentrating participation at the expense of small and medium competitors and reducing the redundancy that protects supply against disruption.

Fourth, under our modeling assumptions, a hypothetical sustained loss of pentavalent and hexavalent supply raises missed doses three- to ten-fold relative to all other scenarios, because lower-valency alternatives lack the near-term capacity to replace combination-vaccine supply at scale. This quantifies, at the procurement-planning level, the coverage cost of replacing combination vaccines with lower-valency alternatives, whatever the cause.

Fifth, we compare three objectives: a balanced coordinating-entity model (UG), a social-benefit model that minimizes missed doses (SB), and a producer profit-maximizing model (MP). UG achieves the lowest administrative cost, SB attains comparable missed doses at greater cost and schedule complexity, and MP performs worst on both. Across all three, the procurement objective, not demand variability, is the dominant driver of cost and schedule complexity, making objective selection the primary policy lever available to planners.

Our results should be interpreted within the context of procurement planning. The model is scoped to tender scheduling within the HCVM framework and does not account for downstream supply chain costs, logistics, vaccine administration, or cold chain infrastructure, and it relies on assumed production capacities that may differ in practice.

\section{Modeling} \label{s:Models}

In this section, we present three variations to the mathematical model designed to address the long-term vaccine tender scheduling problem. The specific model objectives are formulated by comparing a Balance model to its competing extremes, minimizing solely missed doses, and maximizing profit.

\subparagraph{Model Notation}

We consider a planning horizon of length $\mathcal{T}$, covering a set of antigens $\mathcal{A}$ and vaccines $\mathcal{V}$ sourced from a set of pre-qualified producers $\mathcal{P}$, a subset of the WHO pre-qualified list. The cost structure includes producer reservation prices $r_{vpt}$ for producer $p$ and vaccine $v$ at time $t$, holding costs $h_v$ (as a proportion of price) per vaccine, a tender set-up cost $g_t$, a missed-dose penalty $\beta$, and a capacity expansion cost $\gamma_p$, per producer $p$, subject to an allowable expansion rate $\kappa$. The reservation price, $r_{vpt}$, is the minimum price at which producer $p$ is willing to supply vaccine $v$ in period $t$; below this price, the producer would decline to participate in the tender. Quantity discounts are governed by bracket bounds $\phi_{vm}^{L}$ and $\phi_{vm}^{U}$, for vaccine $v$ in discount bracket $m$, with corresponding discount rates $\zeta_{vm}$, which monotonically increase in larger brackets. The parameter $\delta_t$ is a time-discount factor reflecting the time value of money, which converts costs in period $t$ to present value and applies uniformly across all vaccines and producers. It is distinct from the quantity discounts, which are specific to each vaccine–producer pair and determined by the all-unit discount brackets described above. Demand for each antigen $d_{at}$ and producer capacities $s_{pt}$ are given parameters, with $\bar{s}_p$ denoting the base capacity of producer $p$, which is shared among the vaccines that the producers make.

The key structural decisions concern the tender design. The binary variable $F_{at\tau}$ indicates whether a tender covers antigen $a$ from period $t$ through $\tau$, with tenders restricted to a maximum length of $\Delta$ periods. The binary variable $W_{pt\tau}$ captures whether producer $p$ commits to supply over the tender duration; $Y_{pt}$ indicates whether producer $p$ is able to produce a vaccine in their portfolio, $V_p$, in period $t$, and $Q_{vpt\tau m}$ represents the vaccine supply commitment made by producer $p$ for the tender length from $t$ to $\tau$, within discount bracket $m$. Based on their capacity and committed quantity, a capacity expansion decision is represented by the integer variable $L_{pt}$, and the producer then determines how many doses of vaccine $v$ to produce at a given period, $X_{vpt}$. The binary variable $Z_{vptm}$ identifies the active quantity-discount segment for each producer--vaccine--period combination. $I_{vt}$ is the inventory level, $V_{vt}$ is the aggregate number of individual vaccine doses administered at the system level, and $S_{at}$ captures antigen shortages. A complete listing of all sets, parameters, and decision variables is provided in Appendix~\ref{app:A}.

\paragraph{Balance Model Formulation [UG]}

In the Balance (UNICEF-Gavi) model, we formulate the tender scheduling problem from the perspective of coordinating entities, such as UNICEF and Gavi. The model seeks to reduce procurement and tendering costs while ensuring that producers are sufficiently incentivized to participate through adequate return on investment (ROI).

The full model formulation is provided below; however, we organize the discussion around the functional role of each component. Table~\ref{tab:notation} in Appendix~\ref{app:A} summarizes the notation.

The UG model formulation is given as:
\begin{multline}
\min \sum_{t\in \mathcal{T}} \delta_t \Bigg(\sum_{a \in \mathcal{A}} \sum_{(t,\tau) \in \mathcal{F}} g_t F_{at\tau}
+ \sum_{v\in\mathcal{V}_{\mathcal{P}}}\sum_{p\in\mathcal{P}}\sum_{m\in\mathcal{M}}\bar{r}_{vt}(1 -\zeta_{vm}) \sum_{(t,\tau) \in \mathcal{F}}Q_{vpt\tau m} \\
+ \sum_{a \in \mathcal{A}} \beta S_{at} + \sum_{v \in \mathcal{V}} h_v\bar{r}_{vt}I_{vt} \Bigg)
\label{eq:UG_MP}
\end{multline}

{\setlength{\jot}{3pt}%
\let\oldsum\sum\renewcommand{\sum}{\textstyle\oldsum}%
\begin{alignat}{2}
(\tau - t + 1) F_{at\tau} &\leq \sum_{p\in \mathcal{P}_a}\sum_{l=t}^{\tau} Y_{pl},
  &\quad& \forall a\in \mathcal{A}, (t,\tau) \in \mathcal{F} \label{eq:tender_MP}\\
\sum_{\tau:(t,\tau) \in \mathcal{F}} F_{at\tau} &\leq 1,
  && \forall a\in \mathcal{A}, t \in \mathcal{T} \label{eq:overlap_MP}\\
\sum_{\substack{(l,k) \in \mathcal{F} \\ l \le t \le k}} F_{alk} &\geq 1,
  && \forall a\in \mathcal{A}, t \in \mathcal{T} \label{eq:tender_cover_MP}\\
\sum_{a\in \mathcal{A}_p} F_{at\tau} &\geq W_{pt\tau},
  && \forall p \in \mathcal{P}, (t,\tau) \in \mathcal{F} \label{eq:tender_granted_cover_1}\\
F_{at\tau} &\leq \sum_{\substack{p \in \mathcal{P}\\ a \in \mathcal{A}_p}} W_{pt\tau},
  && \forall a \in \mathcal{A}, (t,\tau) \in \mathcal{F} \label{eq:tender_granted_cover_2}\\
\sum_{v \in \mathcal{V}_p} \sum_{m\in \mathcal{M}} Q_{vpt\tau m} &\leq W_{pt\tau} \sum_{l=t}^{\tau}\big[\bar{s}_{p} \nonumber\\
  &\quad + \sum_{k=1}^{l}\kappa(\tau-t+1)\bar{s}_{p} L_{pk}\big],
  && \forall p \in \mathcal{P}, (t,\tau) \in \mathcal{F} \label{eq:commitment_capacity_balance1}\\
&\mathrlap{\sum_{v\in\mathcal{V}_p} \sum_{m\in \mathcal{M}} \bar{r}_{vt}(1-\zeta_{vm}) Q_{vpt\tau m}} \nonumber\\
  &\geq \sum_{v\in\mathcal{V}_p} (1 + l_{vp}) f_{vpt\tau} W_{pt\tau},
  && \forall p\in \mathcal{P}, (t,\tau) \in \mathcal{F} \label{eq:UG_producer_ROI2}\\
\phi_{vm}^{L} Z_{vptm} &\leq \sum_{\tau = t}^{\mathcal{T}_t'} Q_{vpt\tau m},
  && \forall v\in \mathcal{V}, p\in \mathcal{P}, t \in \mathcal{T}, m \in \mathcal{M} \label{eq:discount_lower}\\
\phi_{vm}^{U} Z_{vptm} &\geq \sum_{\tau = t}^{\mathcal{T}_t'} Q_{vpt\tau m},
  && \forall v\in \mathcal{V}, p\in \mathcal{P}, t \in \mathcal{T}, m \in \mathcal{M} \label{eq:discount_upper}\\
\sum_{m \in \mathcal{M}} Z_{vptm} &= 1,
  && \forall v\in \mathcal{V}, p\in \mathcal{P}, t \in \mathcal{T} \label{eq:Z_sum}\\
\sum_{m\in \mathcal{M}} Q_{vpt\tau m} &\geq \sum_{l=t}^{\tau} X_{vpl},
  && \forall v\in \mathcal{V}, p \in \mathcal{P}_v, (t,\tau) \in \mathcal{F} \label{eq:commitment_geq_delivered1}\\
\sum_{v\in \mathcal{V}_p} X_{vpt} &\leq Y_{pt}\big(s_{pt}+\sum_{l=1}^{t}\kappa\bar{s}_{p} L_{pl}\big),
  && \forall p \in \mathcal{P}, t \in \mathcal{T} \label{eq:delivered_leq_capacity1}\\
I_{v,t-1} + \sum_{p\in\mathcal{P}_v} X_{vpt} &= V_{vt} + I_{vt},
  && \forall v \in \mathcal{V}, t \in \mathcal{T} \label{eq:inventory_balance}\\
d_{at} - \sum_{v \in \mathcal{V}_a} V_{vt} + S_{a,t-1} &\leq S_{at},
  && \forall a\in \mathcal{A}, t \in \mathcal{T} \label{eq:missed_dose_balance}\\
I_{v0} &= \hat{I}_{v0}, && \forall v \in \mathcal{V} \label{eq:inv_initial}\\
F_{at\tau}, Y_{pt}, W_{pt\tau}, Z_{vptm} &\in \{0,1\},\quad L_{pt} \in \mathbb{Z}^{+} \label{eq:UG_binary_constraints}\\
Q_{vpt\tau m}, X_{vpt}, I_{vt}, S_{at}, V_{vt} &\geq 0 \label{eq:UG_geq_constraints}
\end{alignat}
}

The objective function \eqref{eq:UG_MP} minimizes the total discounted system cost over the planning horizon. This includes: (i) tendering costs, (ii) procurement costs that account for price discounts based on committed volumes, (iii) penalties associated with unmet demand (missed doses), and (iv) inventory holding costs. Future costs are discounted using factor $\delta_t$ to account for the multi-period nature of the problem.

The constraints can be interpreted in five main groups:

\noindent {\bf Tender scheduling, commitment, and coverage constraints.}
Constraints \eqref{eq:tender_MP}--\eqref{eq:commitment_capacity_balance1} govern the structure and timing of tenders and define the logical relationship between tender awards and producer participation, linking four sets of decision variables. The binary variable $F$ schedules a tender for antigen $a$, from period $t$ to $\tau$, establishing the active window over which procurement occurs. Constraint~\eqref{eq:tender_MP} ensures that a tender for an antigen can be scheduled only if a producer able to supply that antigen has production ($Y = 1$) in every period $t$ covered by the tender ($F$); note that this requires continuous \emph{production} across the tender window, not continuous \emph{delivery} --- a producer can satisfy this constraint by producing every period while still using $W_{pt\tau}$ to commit deliveries only in select periods, pre-building inventory in intervening years to meet the demand of the periods it does commit to. Constraint \eqref{eq:overlap_MP} ensures that, for a given antigen $a$, at most one tender may begin in period $t$, regardless of its duration $\tau$. This does not preclude tenders from overlapping in time; it simply prevents two tenders for the same antigen from being scheduled to start in the same period. Constraint \eqref{eq:tender_cover_MP} enforces rolling coverage: for every period $t$, at least one active tender $F$ must currently cover the demand for every antigen, ensuring continuous coverage throughout the planning horizon without requiring any single tender to span it. The binary variable $W$ indicates whether a producer commits to delivering vaccines in a given period. Constraints~\eqref{eq:tender_granted_cover_1}--\eqref{eq:tender_granted_cover_2} link tenders $F$ and producer commitments $W$ in both directions. Constraint~\eqref{eq:tender_granted_cover_1} ensures that a producer $p$ can commit to a tender starting in period $t$ with duration $\tau$ only if such a tender is scheduled for at least one antigen $a \in \mathcal{A}_p$ that $p$ can supply. Conversely, constraint~\eqref{eq:tender_granted_cover_2} ensures that a tender for antigen $a$ can be scheduled only if at least one producer able to supply $a$ commits to it. The variable $Q$ captures the quantity committed by a producer, while constraint \eqref{eq:commitment_capacity_balance1} ensures that total committed quantities do not exceed available production capacity, reflecting capacity expansion decisions made by producers for $W$ commitments to participate. Together, these four variables encode the full scheduling logic: a tender is opened ($F$), producers are selected ($Y$), commitments are ensured ($W$), and committed quantities are planned ($Q$).

\noindent {\bf Producer incentive (ROI) constraints.}
Constraint \eqref{eq:UG_producer_ROI2} ensures that each producer achieves a minimum return on investment, relative to their committed volume, $Q$. This is achieved by requiring committed quantities, $Q$, to be sufficiently large to recover a portion of production setup costs over time, ensuring cost recovery and encouraging both participation and higher commitment levels when possible.

\noindent{\bf Discount structure constraints.}
Constraints \eqref{eq:discount_lower}--\eqref{eq:Z_sum} model a piecewise discounting scheme, where procurement volumes are assigned to discrete pricing segments. These constraints ensure that committed quantities fall within the appropriate discount thresholds, implementing an all-unit discount setup.

\noindent{\bf Commitment to delivery constraints.}
Planned commitments, $Q$, are then linked via constraint \eqref{eq:commitment_geq_delivered1} to realized deliveries, $X$, ensuring that deliveries do not exceed committed volumes, while $W$ ensures that commitment quantities are only planned for committed times, $t$. Constraint \eqref{eq:delivered_leq_capacity1} enforces that realized deliveries, $X$, in each period remain within a producer's capacity limits, for awarded tenders, $Y$.

\noindent{\bf Inventory and demand balance constraints.}
Constraints \eqref{eq:inventory_balance}--\eqref{eq:missed_dose_balance} track the flow of vaccines and unmet demand over time as standard flow balance constraints. Constraint \eqref{eq:inventory_balance} enforces conservation of vaccine inventory across periods. Constraint \eqref{eq:missed_dose_balance} tracks unmet demand (missed doses), allowing for carryover between periods. Constraint \eqref{eq:missed_dose_balance} is written as an inequality because a single dose of a multivalent vaccine counts toward every antigen it contains, so supplying enough doses to meet demand for one antigen can exceed demand for another antigen in the same vaccine (e.g., penta doses needed to cover diphtheria demand may exceed Hib demand); under equality, this over-coverage would force $S_{at} < 0$ and render the model infeasible, whereas the inequality allows $S_{at} = \max\{0,\, d_{at} - \sum_{v \in \mathcal{V}_a} V_{vt} + S_{a,t-1}\}$ whenever $S_{at}$ is penalized in the objective.

Finally, constraints \eqref{eq:UG_binary_constraints}--\eqref{eq:UG_geq_constraints} define the integrality and non-negativity requirements of the decision variables.

\subparagraph{Social Benefit Model Formulation [SB]}

The objective of this model is to minimize the number of missed doses of vaccines over the tender period.

The SB model formulation is given as
\begin{equation}
   \min \sum_{t \in \mathcal{T}}\sum_{a \in \mathcal{A}} S_{at} \label{OBJ_SB}
\end{equation}
subject to constraints \eqref{eq:tender_MP}--\eqref{eq:UG_geq_constraints}.

\subparagraph{Max Profit Model Formulation [MP]}

The objective of this model is to maximize producers' total revenue through commitments, while considering the cost of capacity expansions and production setup.

The MP model formulation is given as:

\begin{multline}
\min \sum_{t\in \mathcal{T}} \delta_t \Bigg(-\sum_{v\in\mathcal{V}_{\mathcal{P}}}\sum_{p\in\mathcal{P}}\sum_{m\in\mathcal{M}} \bar{r}_{vt}(1 -\zeta_{vm}) \sum_{\tau = t}^{\mathcal{T}}Q_{vpt\tau m} \\
+ \sum_{p \in \mathcal{P}} \gamma_p L_{pt} + \sum_{v\in \mathcal{V}_{p}} \sum_{p \in \mathcal{P}} \sum_{\tau = t}^{\mathcal{T}} f_{vpt\tau}W_{pt\tau}\Bigg)
\label{master_problem}
\end{multline}

subject to constraints \eqref{eq:tender_MP}--\eqref{eq:UG_geq_constraints}.

\paragraph{Linearization of models}
Constraints \eqref{eq:commitment_capacity_balance1} and \eqref{eq:delivered_leq_capacity1} contain bilinear terms in which one factor is binary; we linearize them exactly using McCormick envelopes \citep{mccormick1976computability, Glover1975}, as detailed in Appendix~\ref{app:B}.

\paragraph{{Data collection and analysis}} \label{s:methods.1}

The computational experiments employ a combination of synthetically generated and empirically grounded data. The analysis focuses on the antigens listed in Table~\ref{tab:vaccines_antigens}, spanning four vaccine groupings relevant to LIC and LMIC immunization programs. We use the 2020-2030 Linksbridge Vaccine Almanac data \citep{Linksbridge}, which provides UNICEF demand forecasts \citep{UNICEF_supply} for vaccine-preventable diseases over a ten-year planning horizon in LIC and LMIC settings. 

\begin{table}[!tb]
\small{ 
\centering
\begin{minipage}{\textwidth}
\caption{Vaccines and corresponding antigens considered in the computational study.}
\label{tab:vaccines_antigens}
\begin{tabularx}{\textwidth}{|l|X|}
\hline
\textbf{Vaccine} & \textbf{Antigens} \\
\hline
Hexa & Diphtheria, Tetanus, Pertussis, \textit{Haemophilus influenzae} type b, Polio, Hepatitis B \\
Penta & Diphtheria, Tetanus, Pertussis, \textit{Haemophilus influenzae} type b, Hepatitis B \\
Rotavirus & Rotavirus \\
HPV & Human papillomavirus \\
PCV & Pneumococcal conjugate \\
\hline
\end{tabularx}
\end{minipage}}
\end{table}

Vaccine cost parameters are derived from historical procurement records published by UNICEF \citep{UNICEF_supply}. As dose prices vary across producers, we adopt the average cost per dose for each vaccine as the unit price in the model, serving as a proxy for the market reservation price.

{\bf Supply and Demand} Vaccine procurement tenders are issued at the antigen level. The Linksbridge Vaccine Almanac provides annual demand forecasts per vaccine. We disaggregate this demand by antigen, and for each antigen $a$, annual demand is computed as the aggregate demand across all vaccines $v \in \mathcal{V}_{a}$, where $\mathcal{V}_{a} \subseteq \mathcal{V}$ denotes the subset of vaccines containing antigen $a$. For instance, demand for diphtheria is obtained by summing demand across all vaccines in the formulary that include the diphtheria antigen. Figure~\ref{fig:Hexa_10} in Appendix~\ref{app:C} illustrates the original UNICEF projected antigen demand over the 2020-2030 planning horizon for the six antigens comprising the hexavalent (DTP-HepB-Hib-IPV) vaccine: diphtheria, tetanus, pertussis, hepatitis B, \textit{Haemophilus influenzae} type b (Hib), and polio. These forecasts carry known limitations, discussed in Appendix~\ref{app:C}.

We construct an independent ten-year demand forecast from 2020--2025 data, consistent with the UNICEF projections, and sample thirty demand portfolios around its mean to support sensitivity analysis (Appendix~\ref{app:C}). Producer capacities, which are proprietary, are synthetically estimated from historically observed delivery volumes \citep{UNICEF_supply}.

\section{Analysis and Discussion of Results} \label{s:Analysis}

\noindent {\bf Experimental Setup}
Model runs were conducted on a university high-performance computing cluster. All optimization problems were solved using the Gurobi Optimizer on an Intel Xeon Gold 6130 processor (32 cores, 192 GB RAM). Each instance was solved to a relative optimality gap of 1\%, with a maximum time limit of 72 hours; however, no runs reached this limit. Mean solve times are reported in Table~\ref{tab:solve_times} (Appendix~\ref{app:C}).

To assess model sensitivity, we solve six problem variants that differ in their assumed discount structure, each evaluated under three objective configurations (UG, SB, MP) and replicated across 30 independent problem instances per configuration. Across the 30 instances, demand for each antigen and time period is independently varied, with per-antigen demand totals exhibiting coefficients of variation ranging from approximately 4.9\% to 7.9\% across the 30 trials, while the discount threshold and discount depth (a 10\% price reduction once the quantity threshold is reached) are held fixed across all instances and problems. This design isolates the effect of demand uncertainty within a given discount structure from the effect of changing the discount structure itself. Problem 1 assumes a linear, single-segment cost function with no quantity discount. Problems 2 through 6 introduce a piecewise two- or three-segment cost function, varying the number of segments and which antigens are eligible for the discount: Problem 2 applies the two-segment discount uniformly across all antigens; Problem 3 retains the two-segment structure but restricts the discount to pentavalent and hexavalent vaccines only, with no discount applied to other antigens; Problem 4 uses the two-segment structure with the discount applied to all antigens, removing pentavalent and hexavalent from the available vaccine pool; Problem 5 restricts the discount to PCV, HPV, and rotavirus vaccines only, with no discount applied to other antigens; Problem 6 extends the uniformly applied discount structure to three segments.

Quantity discounts heavily dictate market participation and supplier engagement within procurement frameworks \citep{Monahan1984}. To evaluate their structural impact within the HCVM context, we systematically varied the number of pricing segments across six distinct problem instances. Because formal quantity discounts are not standard practice in UNICEF's currently published pricing, we calibrated the discount breakpoints through a sensitivity analysis over procurement volume (Appendix~\ref{app:C}). The primary breakpoint is set at 300,000 units, where market participation begins to tighten, and the second breakpoint of the three-segment structure at 1,000,000 units, beyond which missed doses show no further improvement.

A similar sensitivity analysis of the missed-dose penalty $\beta$ shows diminishing returns once $\beta$ exceeds \$2.0 (Appendix~\ref{app:C}).

\paragraph{Analytical Measure Definitions}
\label{sec:analytical_measures}

\subparagraph{Coordinating Entity}
\label{sec:coord_ent}

Analysis of tender schedules considers both the schedule, as well as the associated cost of organizing a tender, incurred by UNICEF/Gavi. We define ``cost to tender'' as the cost incurred when organizing the tender and associated procurement decision costs, shown in equation \eqref{eq:coord_business_costs}: 
\begin{multline}
\sum_{t\in \mathcal{T}} \delta_t \Bigg(\sum_{a \in \mathcal{A}} \sum_{\tau \in \mathcal{T}_t'} g_t F_{at\tau} \\
+ \sum_{v\in\mathcal{V}_{\mathcal{P}}}\sum_{p\in\mathcal{P}}\sum_{m\in\mathcal{M}}\bar{r}_{vt}(1 -\zeta_{vm}) \sum_{\tau = t}^{\mathcal{T}_t'}Q_{vpt\tau m} \Bigg)
\label{eq:coord_business_costs}
\end{multline}
We further consider schedule stability as a performance criterion, measured using cosine similarity and its complement, cosine distance, computed per model objective. Originally introduced as a vector-space similarity measure in information retrieval \citep{Salton1987}, cosine similarity has since been adopted broadly in machine learning and optimization to quantify structural agreement between encoded solutions \citep{Gao2024}. Each tender schedule is encoded as a vector $\mathbf{s} \in \mathbb{R}^T$, where $T$ denotes the number of time periods and each element $s_t \in \{0,1\}$ indicates whether procurement occurs in period $t$. Cosine similarity between two schedule vectors $\mathbf{s}_i$ and $\mathbf{s}_j$ is then defined as 
\begin{equation}
    \text{cos\_sim}(\mathbf{s}_i, \mathbf{s}_j) = \frac{\mathbf{s}_i \cdot \mathbf{s}_j}{\|\mathbf{s}_i\| \|\mathbf{s}_j\|},
\end{equation}
capturing structural consistency in tender timing independent of scale \citep{Salton1987}. This measure serves a dual purpose in the analysis: when applied across the 30 problem instances within a single model, it quantifies \textbf{schedule stability} — the degree to which a model repeatedly produces structurally similar tender schedules under demand variation; when applied across aggregate problem instances from different models, it quantifies \textbf{schedule consistency} — the degree to which the three models converge on a common tender schedule structure. A value of one indicates identical tender schedule structure, while a value approaching zero reflects orthogonal, structurally dissimilar tender schedules. Cosine distance, $\text{cos\_dist}(\mathbf{s}_i, \mathbf{s}_j) = 1 - \text{cos\_sim}(\mathbf{s}_i, \mathbf{s}_j)$, serves as the corresponding divergence measure \citep{Gao2024}, enabling direct comparison of tender schedule stability across the UG, SB, and MP objectives.

{\bf Market Participation}
\label{sec:participation} To assess market participation across all models, we consider both the number of producers participating in the market, as well as total revenue generated by each producer in the system, defined as the price of a vaccine multiplied by total vaccine commitments, less any capacity expansion costs and unrecouped production costs in a given time period, as shown in equation \eqref{eq:producer_cost}. This net revenue allows us to assess how producer participation responds when positive (producer-side) and negative (buyer-side) discount levels are introduced. The model does not impose any categorical grouping of producers. The labels small, medium, and large are introduced for descriptive convenience in the presentation of results, reflecting each producer's initial production capacity, assumptions necessary for this model.
\begin{multline}
\sum_{t\in \mathcal{T}} \delta_t \Bigg(\sum_{v\in\mathcal{V}_{\mathcal{P}}}\sum_{p\in\mathcal{P}}\sum_{m\in\mathcal{M}}\bar{r}_{vt}(1 -\zeta_{vm}) \sum_{\tau = t}^{\mathcal{T}_t'}Q_{vpt\tau m} \\
- \sum_{p \in \mathcal{P}} \gamma_p L_{pt} - \sum_{v\in \mathcal{V}_{p}} \sum_{p \in \mathcal{P}} \sum_{\tau = t}^{\mathcal{T}} (1-f_{vpt\tau})W_{pt\tau}\Bigg)
\label{eq:producer_cost}
\end{multline}

\paragraph{Results by Stakeholder Perspective}
%   -- Reorganize results around WHO is affected, not how measures are defined.

\subparagraph{Coordinating Entity}
\label{sec:coord_ent_results}

\begin{figure}[!tb]
  \centering
  \begin{minipage}[t]{0.48\linewidth}
    \centering
    \includegraphics[width=\linewidth]{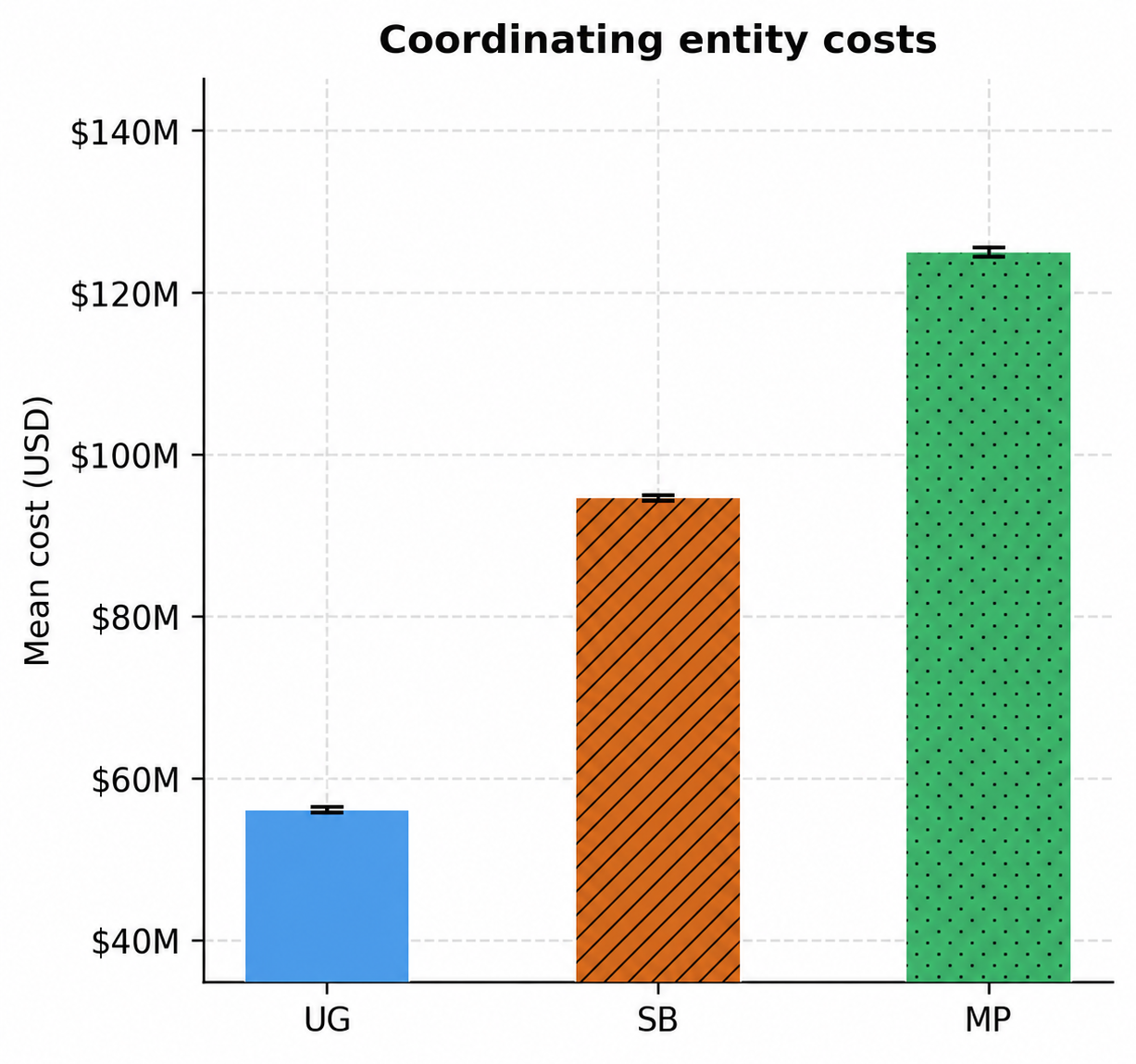}
    \caption{\textbf{Problem 2---mean annual coordinating-entity cost-to-tender by model objective}}
    \label{fig:coord_costs_by_model}
  \end{minipage}\hfill
  \begin{minipage}[t]{0.48\linewidth}
    \centering
    \includegraphics[width=\linewidth]{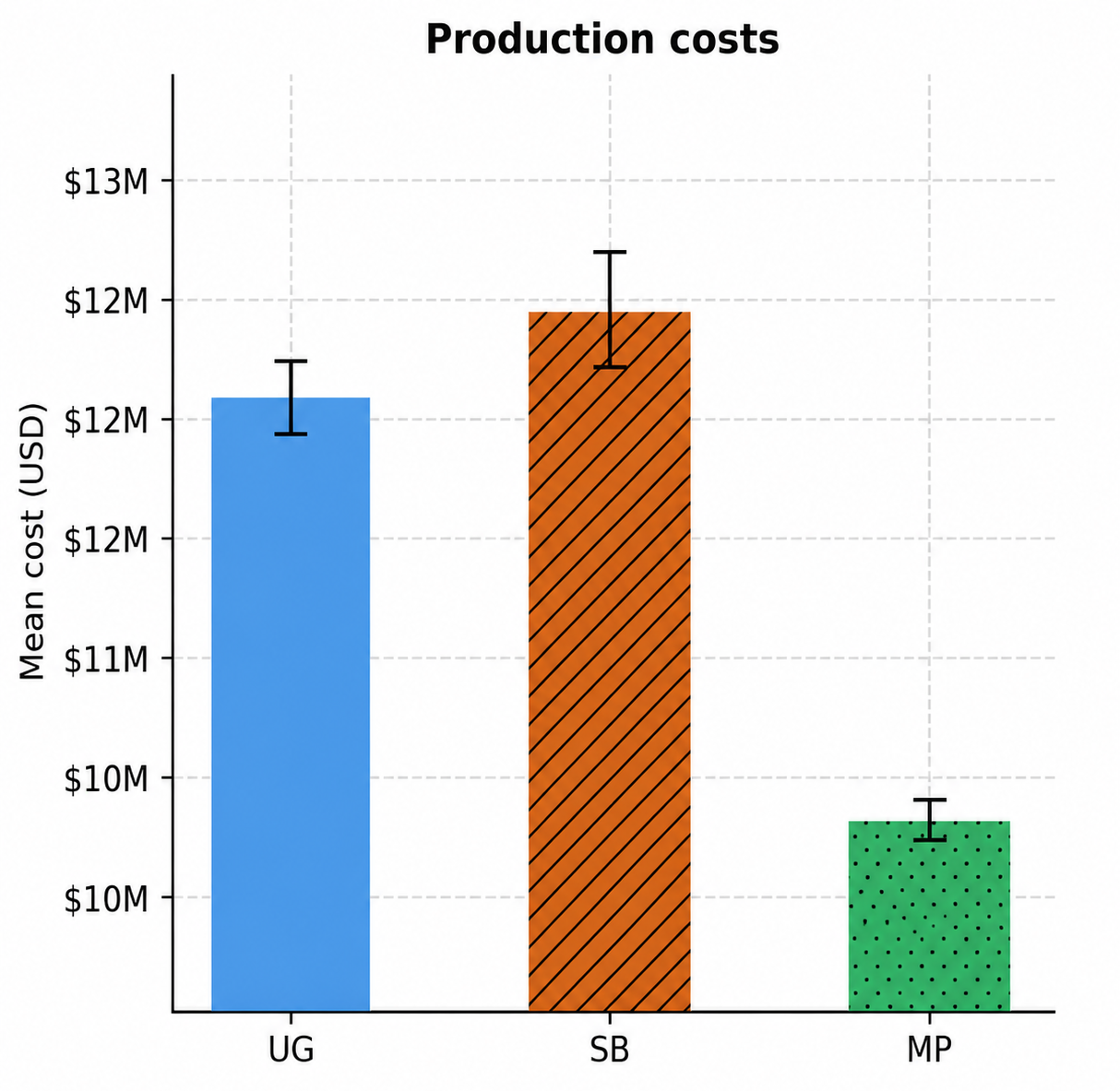}
    \caption{\textbf{Problem 2---mean annual production costs by model objective}}
    \label{fig:Production_costs}
  \end{minipage}
\end{figure}
The cost-to-tender metric serves as a basis for cross-model comparison from the coordinating entity's perspective. As the function in \eqref{eq:coord_business_costs} is only explicitly minimized under the UG model, the remaining objectives do not explicitly contain tendering costs, and as a result, cost-to-tender increases from UG to SB to MP, respectively. While not included in the objective of the SB and MP models, cost-to-tender remains a key metric for coordinating entity planners, and is therefore assessed across all models, regardless of whether or not it is included in respective objective functions. Under the SB model, tendering costs are higher relative to the UG model, as the objective prioritizes coverage maximization. Cost-to-tender is highest under the MP model, wherein producers seek to maximize revenue through commitments of large vaccine quantities. The tendency for larger commitments follows directly from the structure of the MP objective: because the price per dose is determined by committed quantity $Q$, the objective rewards securing new commitments, while the model requires manufacturers to maintain sufficient capacity to fulfill them. Commitments are, however, not unlimited, as constraint \eqref{eq:commitment_capacity_balance1} ensures that commitments do not exceed a producer's available capacity. The Balance model penalizes inventory holding through explicit holding costs; the MP objective does not. This omission runs counter to the logic of perishability, since it removes any disincentive to accumulate stock that may expire before use; however, we do not include holding costs in the Max Profit model, as they are not incurred by producers. Explicitly modeling perishability by product type, including differing storage requirements and their associated capacity constraints, would substantially increase model complexity. To maintain computational tractability, we therefore do not model explicit perishability or product-specific storage requirements in this study. Across all three model objectives, cost-to-tender is therefore ordered as UG $\leq$ SB $\leq$ MP, as shown in Figure~\ref{fig:coord_costs_by_model}.

\begin{figure}[!tb]
    \centering
    \includegraphics[width=0.65\linewidth]{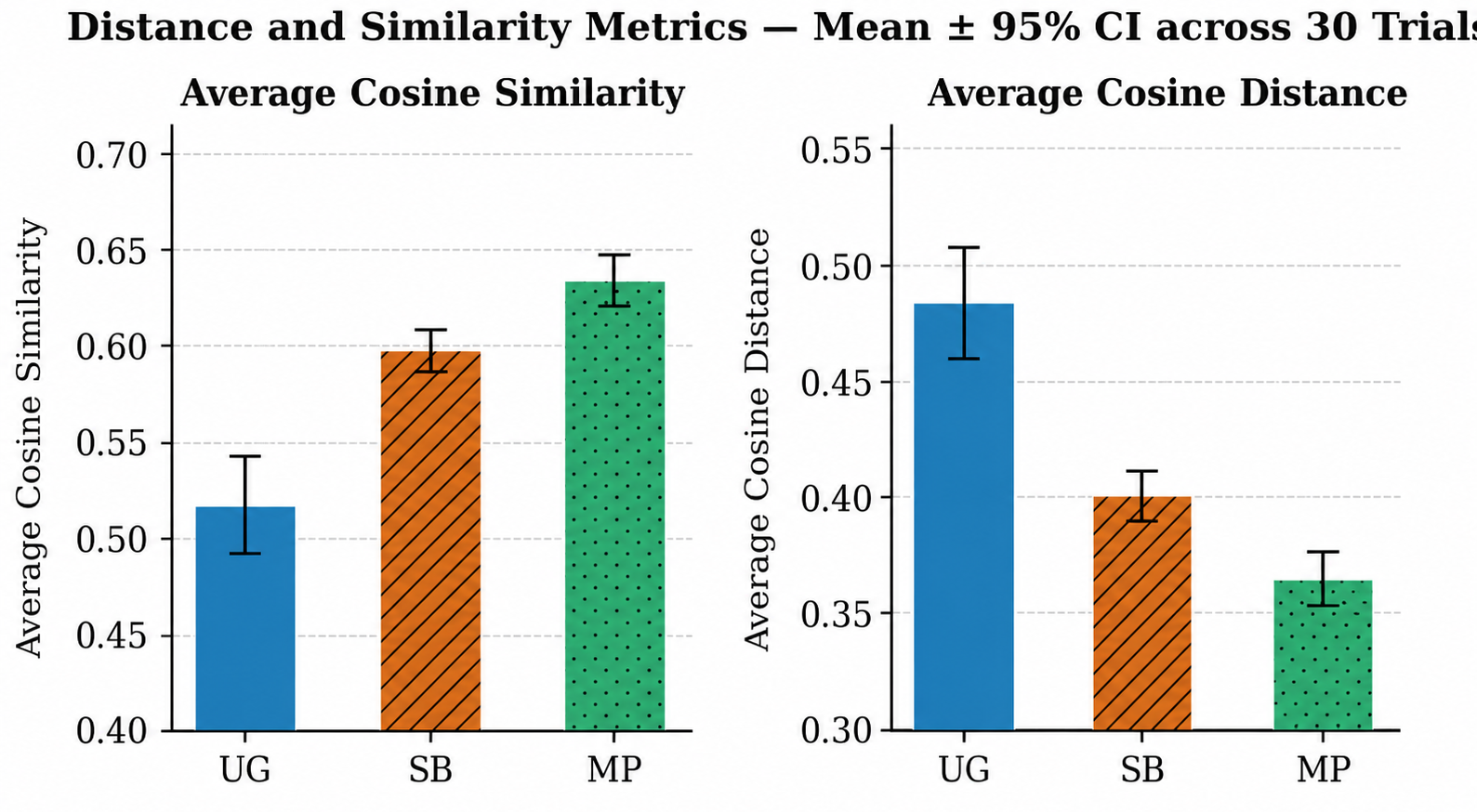}
    \caption{\textbf{Problem 2---tender schedule similarity by model objective}}
    \label{fig:similarity_by_model}
\end{figure}

Similar to coordinating entity costs, a stepwise increase in schedule similarity is observed across models, for all problems (See Figure \ref{fig:similarity_by_model}). As model objectives shift, tender schedules stabilize accordingly. While Problem instance 2 is shown graphically, this trend holds across all problem instances tested, highlighting the part the objective function plays overall. Notably, the UG model produces relatively volatile schedules, as its objective balances the cost of tendering and procuring vaccines against missed doses. This same objective, in conjunction with the penalty for missed doses, also constrains the set of overlapping tenders. By contrast, the SB and MP models introduce additional overlapping tenders to minimize missed doses and enhance profit, respectively, as tender costs are not considered in their objective functions. As the procurement objective shifts away from the balanced UG objective toward the more specialized SB and MP objectives, the number of tenders increases, resulting in greater overlap among procurement periods and, consequently, greater scheduling complexity. Table~\ref{tab:overlaps_by_model} illustrates this relationship, suggesting that policymakers should anticipate increased coordination requirements when pursuing more specialized procurement objectives, where the cost to tender is not the primary objective of the model.

\begin{table}[!tb]
\centering
\caption{Mean number of overlapping tender periods by antigen and objective, Problem 2 ($\pm$ 95\% CI, 30 trials).}
\label{tab:overlaps_by_model}
\small
\begin{tabular}{lccc}
\toprule
\textbf{Antigen} & \textbf{UG} & \textbf{SB} & \textbf{MP} \\
\midrule
Diphtheria  & 2.97 $\pm$ 0.63 & 17.87 $\pm$ 2.66 & 20.63 $\pm$ 2.00 \\
HPV         & 1.33 $\pm$ 0.30 & 3.90 $\pm$ 1.47  & 65.47 $\pm$ 2.08 \\
Hepatitis B & 3.90 $\pm$ 0.78 & 15.73 $\pm$ 2.21 & 15.57 $\pm$ 2.66 \\
Hib         & 5.43 $\pm$ 0.91 & 17.80 $\pm$ 2.27 & 14.77 $\pm$ 2.39 \\
PCV         & 0.60 $\pm$ 0.20 & 3.27 $\pm$ 1.00  & 38.90 $\pm$ 1.58 \\
Pertussis   & 4.47 $\pm$ 0.68 & 18.97 $\pm$ 2.22 & 13.63 $\pm$ 2.11 \\
Polio       & 0.00 $\pm$ 0.00 & 0.03 $\pm$ 0.06  & 0.40 $\pm$ 0.47 \\
Rotavirus   & 0.03 $\pm$ 0.06 & 0.20 $\pm$ 0.14  & 3.37 $\pm$ 1.35 \\
Tetanus     & 3.17 $\pm$ 0.57 & 17.70 $\pm$ 2.60 & 21.87 $\pm$ 2.13 \\
\bottomrule
\end{tabular}
\end{table}

\begin{figure}[!tb]
  \centering
  \begin{minipage}[t]{0.48\linewidth}
    \centering
    \includegraphics[width=\linewidth]{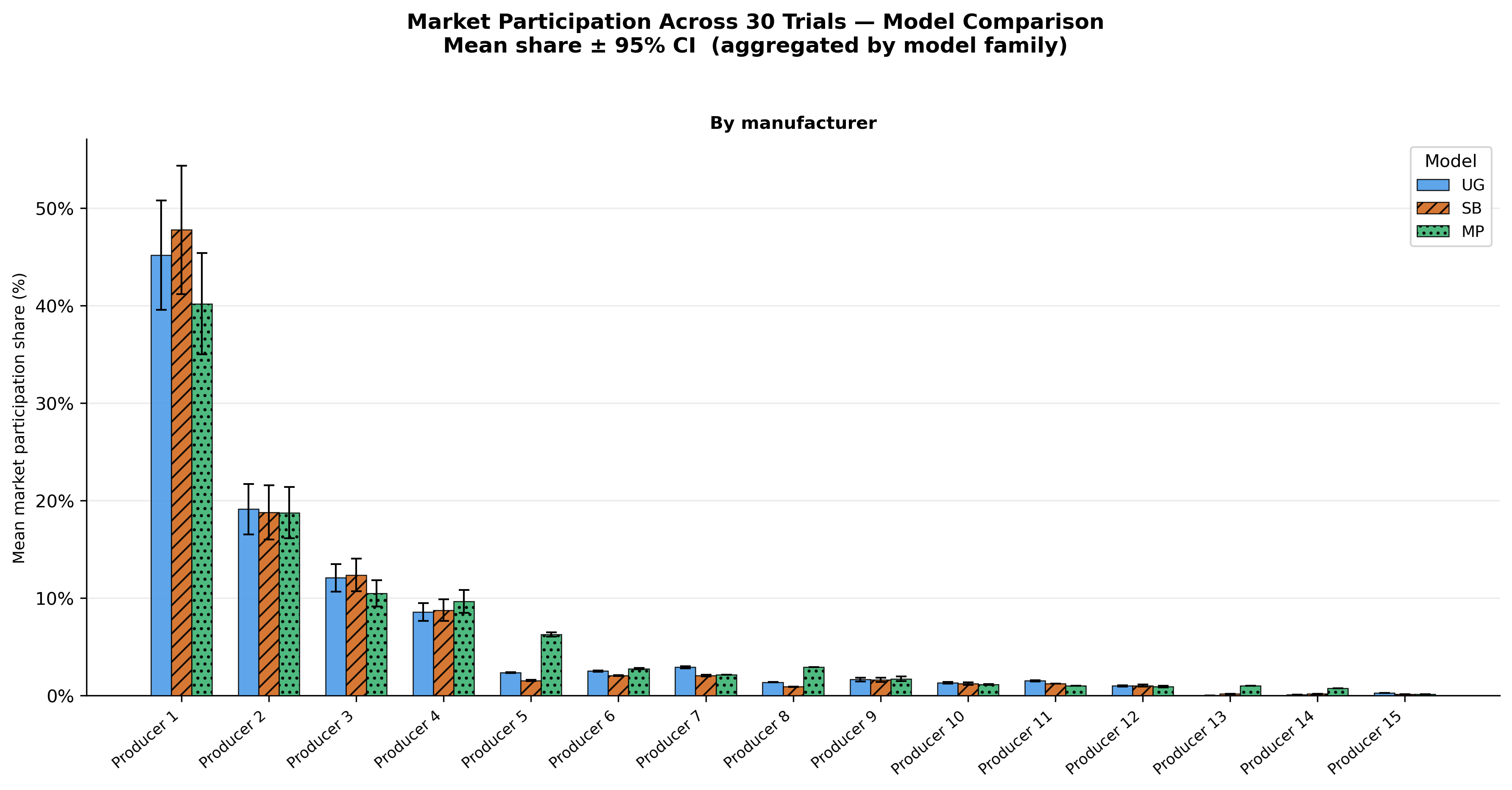}
    \caption{\textbf{Problem 2---market participation by volume: producer}}
    \label{fig:qty_prod}
  \end{minipage}\hfill
  \begin{minipage}[t]{0.48\linewidth}
    \centering
    \includegraphics[width=\linewidth]{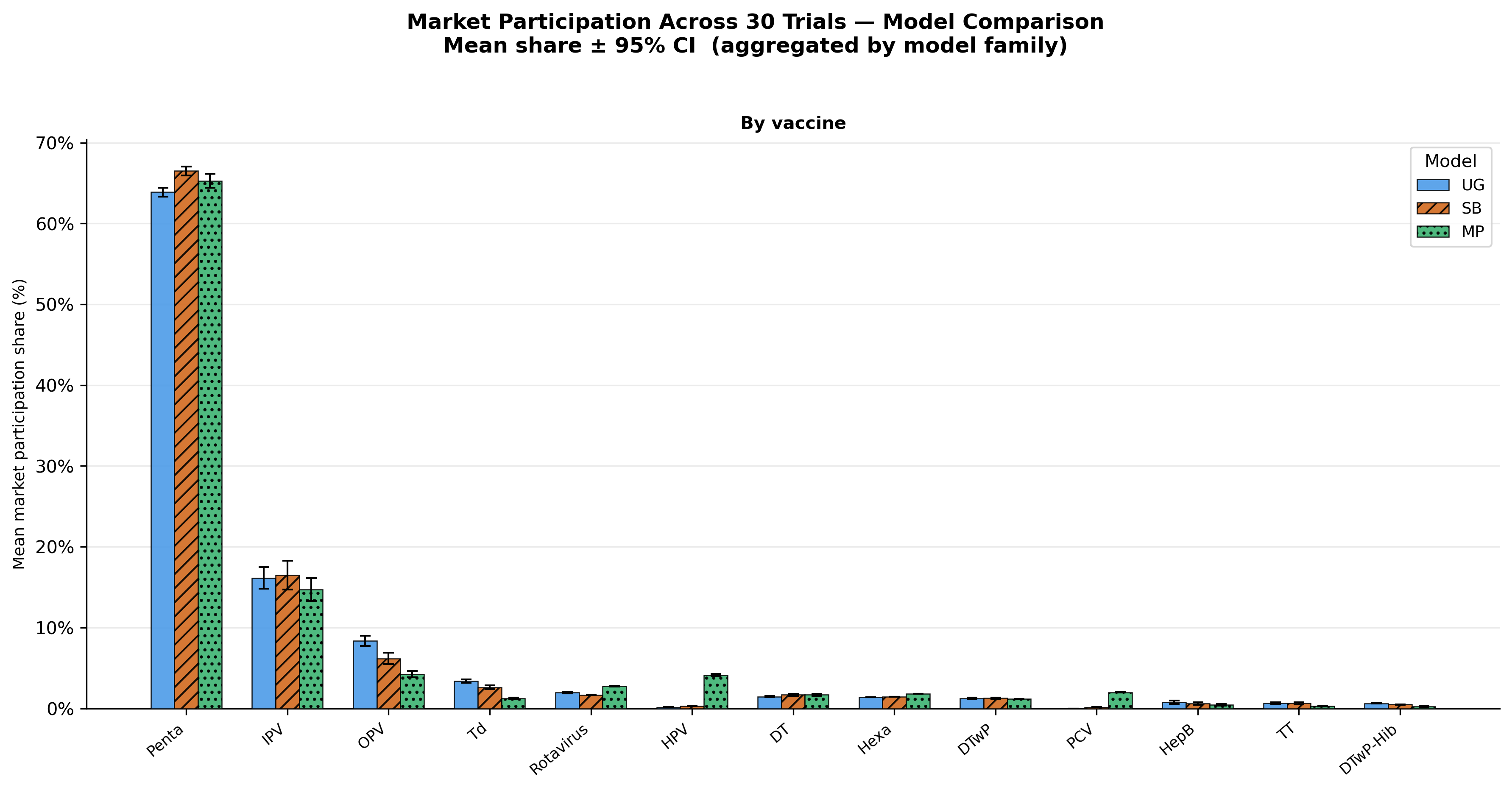}
    \caption{\textbf{Problem 2---market participation by volume: vaccine}}
    \label{fig:participation_by_vaccine}
  \end{minipage}
\end{figure}

{\bf Producers}
%     - Production costs, revenue, and profit by model objective
%     - Market participation and concentration
Figures \ref{fig:qty_prod} and \ref{fig:participation_by_vaccine} present total committed quantities disaggregated by producer and vaccine type, respectively. Pentavalent vaccine dominates procurement across all scenarios in which it is included, accounting for the largest share of total committed doses by a substantial margin, followed by IPV and OPV. This reflects both the breadth of antigen coverage offered by combination vaccines and the correspondingly high demand that follows from their valency. In this hypothetical market, the pool of hexavalent vaccine producers is limited to a single producer, resulting in constrained committed quantities; as the number of qualified hexavalent producers increases, a broader distribution of committed quantities is expected to follow.

At the producer level, market allocation is similarly concentrated: Four large-scale producers collectively account for the majority of committed quantities, with the remaining producers contributing comparatively modest volumes. This is because larger producers are better positioned to absorb high-volume tender commitments, and as a result, market share naturally accrues to those with the greatest production capacity. Notably, production costs are lowest under the MP model (see Figure \ref{fig:Production_costs}). The underlying objective function discourages unnecessary capacity expansion; since additional capacity investment reduces net revenue, producers operating under this objective have less incentive to commit quantities that maximize margin rather than volume. 

{\bf System-Level (Holding Costs)} Holding costs are broadly similar across models and problem instances, with one exception: the MP model shows an increase in holding costs for problem instance 2. Because the MP objective is designed to favor producers, it drives substantially larger committed vaccine quantities. While this benefits coverage, it may also carry an unintended consequence---increased vaccine wastage---that we do not measure directly in this analysis. As discussed previously, the MP objective favors commitments over on-hand inventory, and this tradeoff is reflected in the observed holding costs.

When only pentavalent and hexavalent vaccines are discounted, however, holding costs under the MP model decrease slightly, as the model instead draws on a mix of lower-valency vaccine commitments and on-hand inventory to meet demand. Since inventory holding costs are fixed in our formulation and were not included in the sensitivity analysis, this finding suggests that profit maximization inherently discourages the use of on-hand inventory in favor of larger committed volumes.

\noindent {\bf Scenario Analysis}

{\bf Effect of Vaccine Portfolio} A demand sensitivity analysis over 30 independent trials (per-antigen demand CVs of roughly 5\% and 8\%; antigen-time-level demand varied by 19\% on average) produced statistically insignificant changes in coordinating entity cost. Cost-to-tender is higher under formulations that omit cost from the objective, since they have no incentive to select cheaper feasible solutions, indicating that it is driven by the objective function rather than by demand variation.

When discounts are applied to both pentavalent and hexavalent vaccines, the models consistently prefer Penta, owing to its lower cost and multi-antigen coverage. The preference is strongest in MP, where producers favor larger Penta quantities even with discounts in place. Market structure reinforces it: several large producers supply Penta, whereas Hexa comes from a single producer, limiting competitive pressure. This reflects current supply assumptions rather than a structural disadvantage; planned expansion of hexavalent capacity is expected to narrow the volume gap and eventually exceed current pentavalent volumes, which would likely shift this dynamic in future analyses \citep{UNICEF_Hexa}.

We also test targeted discounts on HPV, rotavirus, and PCV, given their high per-dose prices and observed coverage gaps. In Problem 5, under a uniform 1\% discount on all vaccines and targeted discounts of 1\%, 10\%, and 20\%, HPV and PCV commitments and deliveries remain negligible and essentially unchanged. Rotavirus commitments are similarly negligible at 1\% and 10\% but rise substantially at 20\%, the only configuration in which rotavirus draws on the discounted pricing segment.

The difference stems from price rather than capacity. Rotavirus producers hold no meaningful capacity advantage, but by capacity-weighted average price rotavirus (\$1.32/dose) is about 2.3 and 2.9 times cheaper than PCV (\$2.97/dose) and HPV (\$3.86/dose). A 20\% discount makes rotavirus competitive with the pentavalent and combination vaccines that dominate shared producer capacity, whereas discounted HPV and PCV remain costlier than undiscounted rotavirus. These monovalent vaccines compete with combination vaccines that cover multiple antigens per dose from shared capacity (roughly an eleven-fold efficiency gap between Penta and PCV at a shared producer), and HPV and PCV are the most expensive formulations in their producers' portfolios, with HPV exceeding 30 times a producer's OPV price in one case.

Price therefore limits participation under a cost-sensitive objective. Under SB, which omits procurement cost, HPV and PCV commitments and deliveries rise from near zero to 163,086 and 148,285, respectively, and increase a further 29\% and 4\% under MP, where higher-priced vaccines yield more manufacturer revenue. Discount design can thus improve rotavirus coverage, but improving HPV and PCV coverage under UG requires subsidies or alternative pricing that bring their cost/benefit ratio closer to that of Penta and Hexa.

% Discount breakpoint design also shapes outcomes. A single-tier, no-discount structure yields lower coordinating entity costs and fewer missed doses, and serves as the baseline. Two- and three-segment structures increase missed doses (Figure~\ref{fig:missed_by_discount}) and raise coordinating entity costs by 141\% and 31.6\%, respectively. The two-segment breakpoint is reachable by producers of all sizes, so all concentrate volume to clear it, causing a broad reallocation of committed quantity. In the three-segment structure, the upper threshold acts as a barrier for small and medium producers, who remain in their achievable lower tier; only large producers concentrate volume there, so fewer $Q$ commitments are reallocated and the cost increase is smaller despite the added breakpoint. Efficiency and participation outcomes are thus sensitive to the granularity of discount design.

% \begin{figure}[!tb]
% \centering
% \includegraphics[width=0.5\linewidth]{images/missed_doses_across_discounts.png}
% \caption{\textbf{UG Model - Missed doses across market discount levels}}
% \label{fig:missed_by_discount}
% \end{figure}
%
%\begin{figure}[!tb]
%  \centering
 % \includegraphics[width=0.5\linewidth]{images/missed_doses_across_discounts.png}
%  \caption{\textbf{UG model---missed doses across market discount levels}}
%  \label{fig:missed_by_discount}
%\end{figure}

\begin{wrapfigure}{l}{0.4\textwidth}
    \centering
    \includegraphics[width=0.4\textwidth]{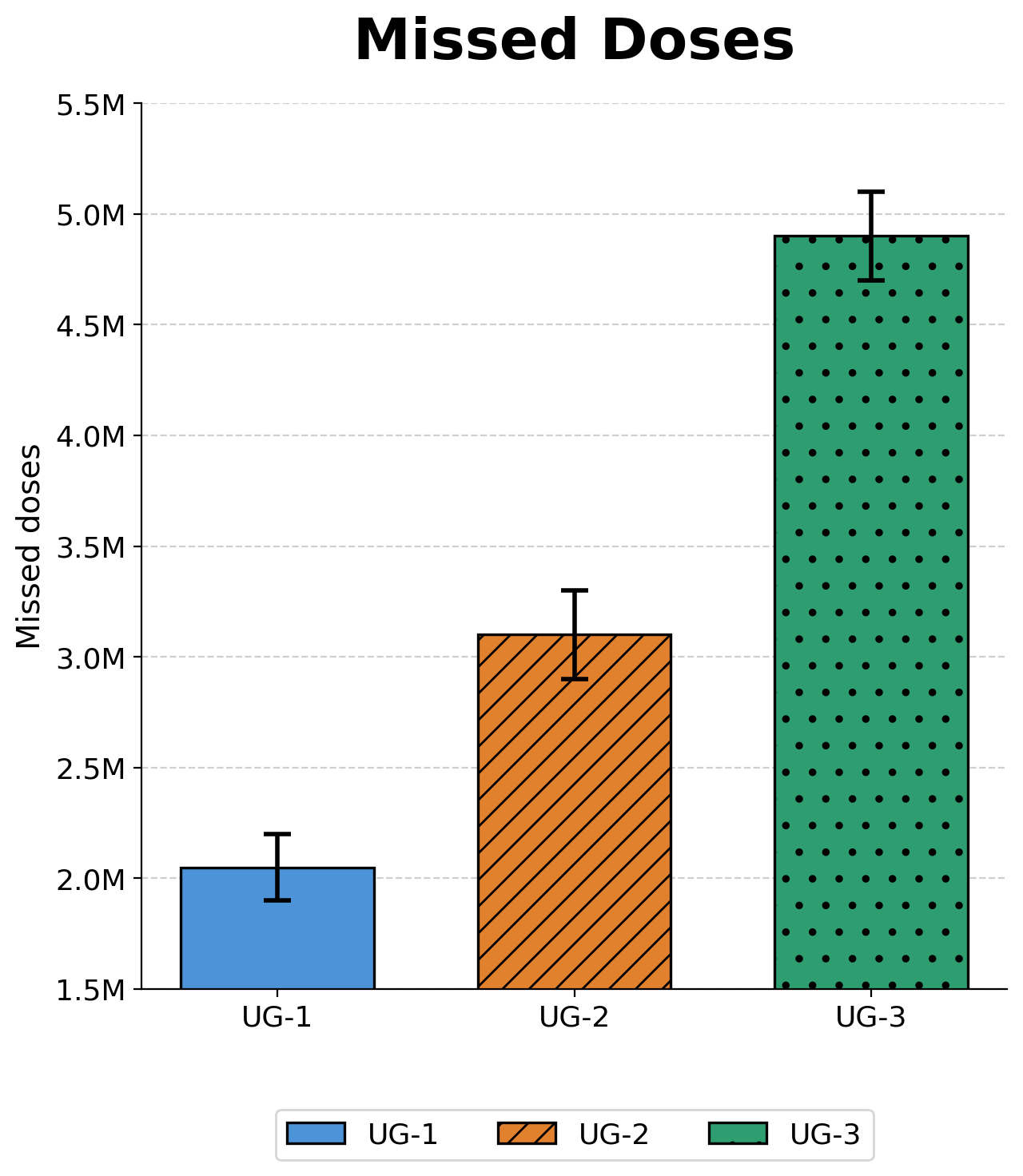}
    \caption{UG model---missed doses across market discount levels}
    \label{fig:missed_by_discount}
\end{wrapfigure}

Discount breakpoint design also shapes outcomes. A single-tier, no-discount structure yields lower coordinating entity costs and fewer missed doses, and serves as the baseline. Two- and three-segment structures increase missed doses (Figure~\ref{fig:missed_by_discount}) and raise coordinating entity costs by 141\% and 31.6\%, respectively. The two-segment breakpoint is reachable by producers of all sizes, so all concentrate volume to clear it, causing a broad reallocation of committed quantity. In the three-segment structure, the upper threshold acts as a barrier for small and medium producers, who remain in their achievable lower tier; only large producers concentrate volume there, so fewer $Q$ commitments are reallocated and the cost increase is smaller despite the added breakpoint. Efficiency and participation outcomes are thus sensitive to the granularity of discount design.

Finally, to simulate a major supply disruption, we remove the highest-valency vaccines, Penta and Hexa, which are the main drivers of coverage, while keeping capacity unchanged. Their antigens must then be covered by lower-valency vaccines, which fragments procurement: more tenders and greater vaccine quantities are needed to meet the same antigen-specific demand. This substantially raises schedule entropy and missed dose rates. The resulting change in coordinating-entity cost therefore reflects a change in procurement composition and structure, not reduced administrative effort due to insufficient vaccine availability.

\subparagraph{Market Participation Patterns}

Across all tested scenarios, sensitivity analysis consistently reveals a market concentrated on large producers, who capture the overwhelming majority of committed dose volume, as shown in Figure \ref{fig:qty_prod}. As producer production capacities are proprietary, we approximate them using real-world contract fulfillment data, under which large producers are assigned higher capacity; this concentration result may therefore partly reflect that assumption rather than an independent model outcome, as larger capacity affords greater opportunity to meet the order volumes required for tiered discounts. This concentration is stable across objective formulations — the choice of procurement objective does not meaningfully alter which producers dominate, only the precise volume each receives. Notably, the distinction between buyer- and producer-applied discount structures is statistically insignificant, with both mechanisms producing nearly identical allocation outcomes, with missed doses rising 1.3\%, and two additional tenders on average. This suggests that producer-side discounts are unfavorable in this model: buyer-side mechanisms achieve, on average, comparable coverage with fewer tenders (51 vs.\ 53), less total procurement (9.18M vs.\ 9.2M doses), and fewer missed doses (2.45M vs.\ 2.49M), indicating buyer-side discounts are the more efficient mechanism at current discount levels, for the intended UNICEF-Gavi market.

Pentavalent vaccine emerges as the dominant structural driver of both market volume and producer concentration, both across objective formulations and discount structures where it is present, as seen in Figure \ref{fig:participation_by_vaccine}. Its presence in the formulary anchors the procurement system around a small set of high-volume producers capable of satisfying multi-antigen demand at scale, while its removal triggers a sharp collapse in total committed volume and a corresponding reorganization of the producer landscape toward a more distributed — but substantially smaller — market. This near-term shortfall is expected to diminish over time as additional manufacturing capacity comes online, new producers enter the pre-qualified market, and competitive pressures rebalance supply and demand across all vaccine valencies. This trade-off between coverage efficiency and market inclusivity represents a fundamental tension that planners must explicitly account for when designing procurement strategy. We again highlight the model assumptions include only one producer of the hexavalent vaccine, therefore we note limited market volume unlike pentavalent vaccine, but expect continued future growth as producers enter the market in the latest UNICEF push for Hexa. 

These findings carry direct implications for coordinating entities. Concentration risk is not an artifact of a particular objective or discount design — it is a structural feature of the market as currently constituted, tied directly to the role of high-valency combination vaccines. Planners should therefore treat formulary composition as a primary lever for managing both coverage outcomes and supply chain risk, and consider whether discount mechanisms can be redirected from cost reduction toward incentivizing broader producer participation in the high-valency segment of the market, a consideration especially germane as the introduction of more hexavalent producers increases.

\paragraph{Overall Effect of Discounts}

We examine buyer-side discounts as a lever for shaping manufacturer
behavior, motivated by the observation that a manufacturer's willingness
to commit capacity is driven largely by expected margin.

Market participation -- whether a manufacturer remains active across the
planning horizon -- proved largely insensitive to segmentation: the large
majority of manufacturers participate regardless of how the market is
segmented, across all instances and objectives. Segmentation and discount
design instead affect \emph{how much} each manufacturer commits, not
\emph{whether} it participates. Table~\ref{tab:scenario_summary_all_models}
reports aggregate commitment ($Q$) by scenario for all three buyer
objectives.

Targeted discounts increased commitment: moving from the balanced UG
objective to the SB objective raised aggregate commitment in every
scenario, most notably in \texttt{2 segments} (43.7M~$\rightarrow$~71.4M
doses, +63\%) and \texttt{Penta/Hexa} (41.3M~$\rightarrow$~66.2M doses).
This gain under SB and MP is partly a structural artifact of the
objective composition, however: holding cost is not penalized, so larger
commitments are not offset by the cost of holding the resulting
inventory.

\begin{table}[!tb]
\centering
\caption{Aggregate commitment (Q) in millions of doses by objective/problem.}
\small{ 
\label{tab:scenario_summary_all_models}
\begin{tabular}{lrrrrrr}
\toprule
 & \multicolumn{2}{c}{\textbf{UG}} & \multicolumn{2}{c}{\textbf{SB}} & \multicolumn{2}{c}{\textbf{MP}} \\
\cmidrule(lr){2-3} \cmidrule(lr){4-5} \cmidrule(lr){6-7}
\textbf{Problem} & \textbf{Mean} & \textbf{SD} & \textbf{Mean} & \textbf{SD} & \textbf{Mean} & \textbf{SD} \\
\midrule
1: 1 segment                         & 11.23 & 0.52 & 15.86 & 2.36 & 130.27 & 0.38 \\
2: 2 segments                        & 43.71 & 1.62 & 71.37 & 1.58 & 83.94  & 0.60 \\
3: 2 segments, penta/hexa discount   & 41.28 & 0.70 & 66.15 & 0.91 & 77.65  & 0.10 \\
4: 2 segments, no penta/hexa         & 8.85  & 0.40 & 9.01  & 0.30 & 15.99  & 0.35 \\
5: 2 segments, HPV/PCV/Rota discount & 43.32 & 0.49 & 51.40 & 0.85 & 51.40  & 0.85 \\
6: 3 segments                        & 40.21 & 2.23 & 43.30 & 1.07 & 62.49  & 0.08 \\
\bottomrule
\end{tabular}}
\end{table}

Two scenario variants isolate the role of combination vaccines and of
product-specific pricing more directly. In the \texttt{no Penta/Hexa}
scenario, the Penta and Hexa combination products are removed from the
manufacturer pool entirely, leaving only monovalent and lower-valency
alternatives to cover the same antigens. This removal is costly on both
sides of the market: aggregate commitment collapses to roughly a fifth of
its level in any other scenario (8.9M--16.0M doses across the three buyer
objectives, versus 40M+ doses when Penta/Hexa are available), and missed
doses rise to their highest level in the study (105M--108M), three to ten
times higher than any other scenario. This suggests that manufacturer
willingness to commit capacity is strongly tied to the availability of
combination-vaccine bundles, and that removing them does not simply shift
volume onto monovalent substitutes -- it suppresses total commitment and
leaves the bulk of demand unserved.

The \texttt{HPV/PCV/Rota} scenario tests a different lever: rather than
removing combination products, it targets Rotavirus (alongside HPV and
PCV) with a discount structure that brings its price component in line
with that of the multi-valent combination vaccines. Under this pricing,
aggregate commitment recovers substantially relative to the
\texttt{no Penta/Hexa} case -- 43.3M--51.4M doses across the three
objectives, four to five times higher -- despite Rotavirus remaining a
single-antigen product. This indicates that price parity with combination
vaccines, applied specifically to Rotavirus, is enough to draw
manufacturer commitment back to levels comparable with the fully bundled
scenarios, without requiring Penta/Hexa themselves to be reintroduced.

Notably, this recovery in commitment volume does not fully translate into
recovered coverage: missed doses in the \texttt{HPV/PCV/Rota} scenario
remain nearly as high as in the \texttt{no Penta/Hexa} case (91M--91.5M,
versus 105M--108M) even though committed volume is four to five times
larger. Taken together, these two scenarios suggest that targeted pricing
can restore manufacturer commitment lost from de-bundling combination
vaccines, but commitment volume alone is not a sufficient proxy for
coverage -- a large share of demand went unmet in both cases, which points
toward the effective coverage per dose of high-valency combination vaccines. 

\paragraph{Implications for Vaccine Procurement Planning}

Our results suggest four implications for coordinating entities planning long-term tenders.

\textbf{Objective selection is the primary lever.} Holding discount structure fixed, cost-to-tender is driven by the procurement objective rather than by demand variation, which produced statistically insignificant cost differences. The balanced UG objective attains missed doses close to those of SB at lower tendering cost and schedule complexity, whereas SB and MP raise both.

\textbf{Discount leverage depends on market maturity, not valency.} Pentavalent vaccine is supplied by a mature, multi-producer market already near its price floor, and hexavalent is expected to follow. Among the monovalent vaccines tested, only rotavirus responded, and only at a 20\% discount; HPV and PCV remained flat across 1--20\% and are likely to require deeper concessions, subsidies, or non-price incentives. Discount design also carries a system-level cost: two- and three-segment structures increased missed doses and raised coordinating-entity costs by up to 141\% over the no-discount baseline. Because producer capacities are synthetic, these thresholds are illustrative rather than calibrated negotiating targets.

\textbf{Volume discounts concentrate rather than broaden the supplier base.} Discounts work by drawing on spare capacity, which mainly large producers hold. Broadening participation is therefore likely to require sustained investment in smaller producers' capacity rather than discounts alone.

\textbf{High-valency vaccine access is a coverage priority.} Removing pentavalent and hexavalent vaccines lowers coordinating-entity cost but sharply increases missed doses and schedule entropy. Because this scenario is a steady-state removal at the market level rather than a simulated disruption, it supports protecting combination-vaccine access but is not a tested model of supply shocks.

\section{Conclusion}\label{s:Conclusion}

This study presents a long-term vaccine tender scheduling framework within the 
Hypothetically Coordinated Vaccine Market, extending current operational planning horizons and incorporating quantity-based discount structures across three competing model objectives. Through sensitivity analysis across demand variability, discount configurations, and vaccine portfolio composition, we derive actionable insights for coordinating entities and procurement planners operating in LIC/LMIC market segments.

Our results demonstrate that the procurement objective is the dominant driver of 
cost-to-tender and schedule complexity, with demand variability playing an insignificant role within each objective. The UG model achieves a pragmatic middle ground — delivering coverage outcomes comparable to the SB model at lower administrative cost and with faster solve times — making it well-suited as a baseline for long-term planning and sensitivity analysis. While the SB model achieves marginal reductions in missed doses, it does so at the cost of increased tender complexity and coordinating entity costs, a trade-off that coordinating entities must weigh against their operational priorities.

With respect to discount design, quantity-based structures offer meaningful leverage only where a vaccine's starting price is low enough for a discount to close the remaining gap to multivalent competitors -- as observed for rotavirus, whose comparatively low price allowed a sufficiently large discount to bring it into real competition with pentavalent and other combination vaccines for shared producer capacity. For higher-priced monovalent vaccines such as HPV and PCV, by contrast, the price gap is too large for any discount rate tested to close: even substantial discounts leave these formulations priced well above their multivalent competitors, and committed volume remains negligible regardless of discount magnitude. Discount mechanisms -- whether applied on the buyer or producer side -- also consistently concentrate market participation among large producers, increasing supply chain vulnerability. Coordinating entities should therefore treat discount design as a viable lever for monovalent vaccines nearing price parity with multivalent alternatives, but not as a substitute for addressing the structural price gap that excludes higher-priced formulations from meaningful market participation. This price-to-valency disadvantage, under a purely cost-minimizing objective, prevents inclusion of the highest-priced monovalent vaccines in large quantities regardless of discount design.

The analysis of vaccine portfolio composition reinforces the critical role of pentavalent and hexavalent vaccines in sustaining coverage. Their removal from the procurement pool produces sharp increases in missed doses and schedule entropy, reflecting the system's dependence on high-valency combination vaccines to efficiently satisfy multi-antigen demand. While near-term shortfalls may diminish as new manufacturing capacity comes online and additional producers enter the pre-qualified market, planners should treat access to high-valency formulations as a foundational requirement for system performance.

Taken together, these findings highlight the sensitivity of vaccine procurement outcomes to objective formulation, discount design, and vaccine portfolio composition, and underscore the value of long-term, model-informed planning as a complement to the iterative, rolling tender frameworks currently employed by organizations such as UNICEF and Gavi. Future work should extend this framework to incorporate downstream supply chain costs, cold chain logistics, and country-level heterogeneity in demand and willingness to pay, as well as explore dynamic discount mechanisms designed explicitly to promote market entry among smaller producers. Additionally, incorporating stochastic programming extensions to more formally address demand and supply uncertainty over extended planning horizons represents a natural and promising direction for further research.

\if0\blind{
\section*{Acknowledgments}

Funding: This work was partially supported by The Bill \& Melinda Gates Foundation, Seattle, WA [grant number INV-044813].

This research is supported by the Arkansas High-Performance Computing Center, which is
funded through multiple National Science Foundation grants and the Arkansas Economic
Development Commission.
} \fi

\section*{Disclosure Statement}
The authors report there are no competing interests to declare.

% Taylor & Francis AI policy (checklist item 11): if generative AI tools were used
% in preparing this manuscript, state that use here, e.g.
\section*{Use of Generative AI}
The authors used Claude Opus 5.5 for grammatical editing, and submission compliance checks; all output was reviewed and edited by the authors.

\FloatBarrier
{\spacingset{1}
\setlength{\bibsep}{2pt plus 1pt}
\bibliographystyle{apacite}
\bibliography{IISE-Trans}}

\clearpage
\appendix
\renewcommand{\theequation}{\thesection.\arabic{equation}}
\setcounter{equation}{0}

\newpage
\section{Model Notation}
\label{app:A}

The complete formulation of the UG model is given in Section~\ref{s:Models}, objective \eqref{eq:UG_MP} and constraints \eqref{eq:tender_MP}--\eqref{eq:UG_geq_constraints}. The SB and MP models replace the objective with \eqref{OBJ_SB} and \eqref{master_problem}, respectively, and retain the same constraint set. Table~\ref{tab:notation} lists all sets, parameters, and decision variables.

{\spacingset{1}\small
\begin{longtable}{@{}p{0.14\linewidth}p{0.82\linewidth}@{}}
\caption{Sets, parameters, and decision variables.}\label{tab:notation}\\
\toprule
\endfirsthead
\toprule
\endhead
\bottomrule
\endlastfoot
\multicolumn{2}{@{}l}{\textbf{Sets}}\\
$\mathcal{A}$ & Antigens \\
$\mathcal{V}$ & Vaccines (from the WHO prequalified vaccine list) \\
$\mathcal{V}_a$ & Vaccines containing antigen $a$ \\
$\mathcal{V}_p$ & Vaccines in producer $p$'s portfolio \\
$\mathcal{A}_v$, $\mathcal{A}_p$ & Antigens in vaccine $v$; antigens in producer $p$'s portfolio \\
$\mathcal{P}$, $\mathcal{P}_v$ & Producers; producers of vaccine $v$ \\
$\mathcal{T}$ & Time periods in the planning horizon \\
$\mathcal{M}$ & Quantity-discount segments \\
$\mathcal{F}$ & Valid tender start/end pairs, $\mathcal{F} = \{(t, \tau) : t \in \mathcal{T},\ t \leq \tau \leq \min\{t + \Delta, |\mathcal{T}|\}\}$ \\
\midrule
\multicolumn{2}{@{}l}{\textbf{Parameters}}\\
$d_{at}$ & Demand for antigen $a$ in period $t$ \\
$s_{pt}$, $\bar{s}_{p}$ & Production capacity of producer $p$ in period $t$; base capacity of producer $p$ \\
$r_{vpt}$, $\bar{r}_{vt}$ & Reservation price of vaccine $v$ from producer $p$ in period $t$; average price of $v$ across producers \\
$l_{vp}$ & Return on investment required by producer $p$ for vaccine $v$ \\
$g_t$ & Set-up cost of initiating a tender in period $t$ \\
$f_{vpt\tau}$ & Proportion of producer $p$'s set-up cost for $v$ over periods $t$--$\tau$ recaptured in the LIC/LMIC segment \\
$h_v$ & Annual holding cost of vaccine $v$ as a proportion of its price \\
$\beta$ & Missed-dose penalty per unit \\
$\gamma_p$, $\kappa$ & Capacity expansion cost of producer $p$; allowable capacity expansion rate \\
$\delta_t$ & Time-discount factor \\
$\Delta$ & Maximum tender length (1--5 periods) \\
$\hat{I}_{v0}$ & Initial inventory of vaccine $v$ \\
$\zeta_{vm}$ & Quantity-discount rate for vaccine $v$ in segment $m$ \\
$\phi_{vm}^{L}$, $\phi_{vm}^{U}$ & Lower and upper quantity bounds of segment $m$ for vaccine $v$ \\
\midrule
\multicolumn{2}{@{}l}{\textbf{Decision variables}}\\
$F_{at\tau}$ & 1 if a tender covers antigen $a$ for periods $t$ through $\tau$; 0 otherwise \\
$Y_{pt}$ & 1 if producer $p$ produces in period $t$; 0 otherwise \\
$W_{pt\tau}$ & 1 if producer $p$ commits to deliver during periods $t$ through $\tau$; 0 otherwise \\
$Z_{vptm}$ & 1 if producer $p$'s commitment of vaccine $v$ in period $t$ falls in segment $m$; 0 otherwise \\
$L_{pt}$ & Capacity expansion of producer $p$ in period $t$ (integer) \\
$Q_{vpt\tau m}$ & Commitment of vaccine $v$ by producer $p$ for periods $t$ through $\tau$ in segment $m$ \\
$X_{vpt}$ & Doses of vaccine $v$ delivered by producer $p$ in period $t$ \\
$I_{vt}$ & Inventory of vaccine $v$ at the beginning of period $t$ \\
$V_{vt}$ & Doses of vaccine $v$ administered in period $t$ \\
$S_{at}$ & Missed doses of antigen $a$ in period $t$ \\
\end{longtable}}

\section{McCormick Linearization}
\label{app:B}
\setcounter{equation}{0}

In our models, constraints \eqref{eq:commitment_capacity_balance1} and \eqref{eq:delivered_leq_capacity1} contain non-convex bi-linear terms. We linearize them using McCormick envelopes \citep{mccormick1976computability}. Since one of the variables in these terms is a binary variable, this substitution yields an
\emph{exact} reformulation of the original bilinear constraint, rather than a relaxation \citep{Glover1975, mccormick1976computability}. Specifically, each bilinear term $xy$ is replaced by an auxiliary variable $\langle xy \rangle^{M}$, representing the linearized form of the original product. The resulting constraints are given below.

\begin{equation}
\langle xy \rangle^M \! \equiv \!
\begin{cases}
    \langle xy \rangle^M \geq x^l y + x y^l - x^l y^l  \\
    \langle xy \rangle^M \geq x^u y + x y^u - x^u y^u  \\
    \langle xy \rangle^M \leq x^l y + x y^u - x^l y^u  \\
    \langle xy \rangle^M \leq x^u y + x y^l - x^u y^l
\end{cases}
\label{eq:McCormick_envelope}
\end{equation}

\paragraph{Constraint \eqref{eq:commitment_capacity_balance1}.} Define
$\hat{L}_{pt\tau} = \sum_{l=t+1}^{\tau}(\tau - l + 1)\kappa\bar{s}_p L_{pl}$ and
$\check{L}_{pt\tau} = \sum_{l=1}^{t}(\tau - t + 1)\kappa\bar{s}_p L_{pl}$, with auxiliary variables $\hat{K}_{pt\tau} = W_{pt\tau}\hat{L}_{pt\tau}$ and $\check{K}_{pt\tau} = W_{pt\tau}\check{L}_{pt\tau}$. Since $W_{pt\tau}\in\{0,1\}$, for all $p\in\mathcal{P}$, $(t,\tau)\in\mathcal{F}$:
{\spacingset{1}
\begin{align}
&\hat{K}_{pt\tau} \geq \hat{L}_{pt\tau} - \hat{L}_{pt\tau}^u (1-W_{pt\tau}), \quad
 \hat{K}_{pt\tau} \leq \hat{L}_{pt\tau}^u W_{pt\tau}, \quad
 \hat{K}_{pt\tau} \leq \hat{L}_{pt\tau}, \\
&\check{K}_{pt\tau} \geq \check{L}_{pt\tau} - \check{L}_{pt\tau}^u (1-W_{pt\tau}), \quad
 \check{K}_{pt\tau} \leq \check{L}_{pt\tau}^u W_{pt\tau}, \quad
 \check{K}_{pt\tau} \leq \check{L}_{pt\tau}, \\
&\textstyle\sum_{v \in \mathcal{V}_p}\sum_{m\in\mathcal{M}} Q_{vpt\tau m} \leq W_{pt\tau}\sum_{l=t}^{\tau}\bar{s}_p + \hat{K}_{pt\tau} + \check{K}_{pt\tau}.
\end{align}}

\paragraph{Constraint \eqref{eq:delivered_leq_capacity1}.} Define $\ddot{L}_{pt} = \sum_{l=1}^{t} \kappa \bar{s}_p L_{pl}$ and $\ddot{K}_{pt} = Y_{pt}\ddot{L}_{pt}$. For all $p\in\mathcal{P}$, $t\in\mathcal{T}$:
{\spacingset{1}
\begin{align}
&\ddot{K}_{pt} \geq \ddot{L}_{pt} - \ddot{L}_{pt}^u (1-Y_{pt}), \quad
 \ddot{K}_{pt} \leq \ddot{L}_{pt}^u Y_{pt}, \quad
 \ddot{K}_{pt} \leq \ddot{L}_{pt}, \\
&\textstyle\sum_{v\in\mathcal{V}_p} X_{vpt} \leq Y_{pt}s_{pt} + \ddot{K}_{pt}.
\end{align}}

\section{Data, Calibration, and Computational Details}
\label{app:C}
\setcounter{equation}{0}

\paragraph{Demand data.} We note that UNICEF and Gavi demand forecasts have historically been both systematically optimistic and highly variable, owing largely to uncertainty in the timing of vaccine introductions \citep{Zuber2009, Dalberg2023PCV}, and the forecasts used here inherit these limitations. We do not aim to forecast the demand signal with high fidelity; rather, we use this data to examine long-term scheduling decisions and to draw broader takeaways about how such scheduling can be approached under demand uncertainty. Similarly, the most accurate demand forecasts would be constructed at the country level, estimating antigen demand directly from projected birth cohort sizes combined with the number of doses of each antigen recommended by national immunization schedules to achieve full coverage. Our disaggregation approach, which relies on vaccine-level demand forecasts and formulary composition, may therefore risk underestimating true antigen demand, since it is bounded by the antigen combinations available in existing vaccine formulations rather than by the full complement of doses recommended for complete immunization.

\begin{figure}[!tb]
    \centering
    \includegraphics[width=0.75\linewidth]{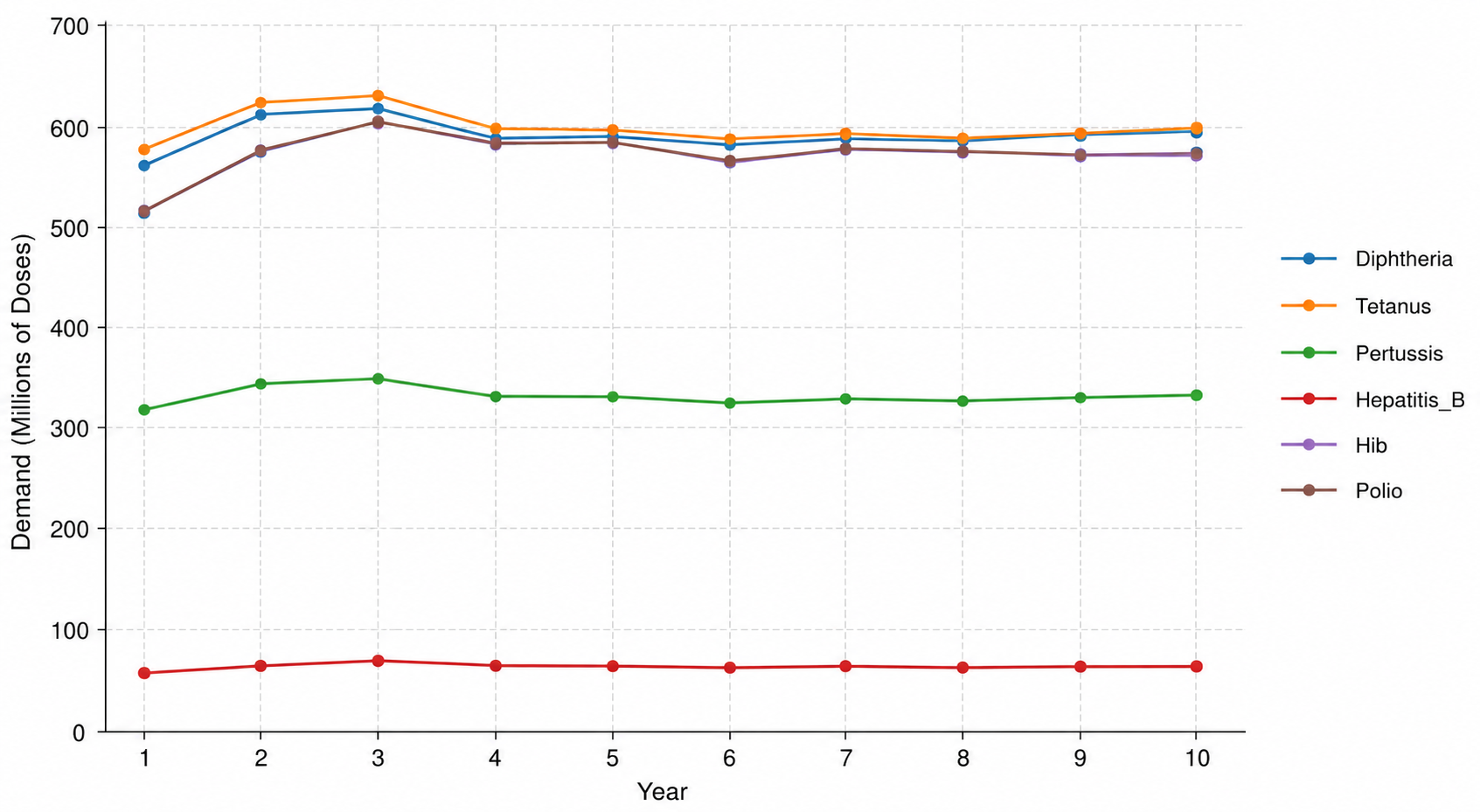}
    \caption{\textbf{UNICEF Projected demand for the six antigens of the hexavalent vaccine over a ten-year planning horizon.}  Adapted from \citet{Linksbridge}.}
    \label{fig:Hexa_10}
\end{figure}

\paragraph{Demand forecast and capacities.} An independent ten-year demand forecast is constructed from the  2020–2025 data, the results of which are  consistent with the aforementioned UNICEF demand projections. Forecasts were generated using a recency-weighted moving average with damped linear trend extrapolation, consistent with the broader literature on exponentially weighted moving-average and damped-trend forecasting methods \citep{Holt2004, McKenzie2010}. The forecasting procedure was tested on historical data and benchmarked against naïve persistence and moving averages; the approach prioritized stable, realistic long-horizon capacity projections over statistical optimality. The forecast mean serves as the distributional center from which thirty demand portfolios are sampled via random normal draws to support systematic sensitivity analysis. Producers' production capacities, which are frequently proprietary and not publicly disclosed, are synthetically estimated for this analysis. Capacity parameters are constructed to reflect historically observed delivery volumes on a per-producer basis \citep{UNICEF_supply}. Furthermore, aggregate production capacity is assumed to meet or exceed total antigen demand up to period $t$ and to scale with forecasted demand over the planning horizon.

\paragraph{Discount breakpoints.} Because formal quantity discounts are not standard practice in UNICEF’s currently published pricing, we conducted a structured sensitivity analysis over breakpoint configurations to empirically determine appropriate discount-pricing boundaries. Discount schedules were modeled as all-unit piecewise functions over procurement volume ($Q$), evaluated across logarithmically spaced grids spanning multiple orders of magnitude. While an initial baseline breakpoint of 100 units successfully induced market participation, it proved operationally unfeasible as standard shipments naturally exceed this floor, rendering the discounts trivial. Instead, a primary breakpoint of 300,000 units was selected to capture where market participation begins to tighten, beyond which a 53\% commitment decrease is seen. In the three-segment model, the second breakpoint is placed at 1,000,000 units — the volume at which the sensitivity analysis indicates missed doses reach a stable lower bound, with no systematic improvement observed at higher procurement thresholds. Beyond 1,000,000 units, the second breakpoint becomes non-binding for most producers, yielding no further increase in commitment — consistent with the breakpoint exceeding the feasible supply volume of producers.  These demonstrate where discrete capacity bounds force a severe trade-off between volume commitments ($Q$) and systemic unfulfilled dose penalties ($S$). While these breakpoints and the associated analysis reflect configurations that were effective within our model, we emphasize that production capacities are synthetically generated and therefore serve primarily to demonstrate the application and structural effects of the breakpoints rather than to provide definitive real-world capacity thresholds.

\paragraph{Missed-dose penalty.} To assess the robustness of the model's missed dose penalty, a sensitivity analysis was conducted on the parameter $\beta$, the penalty assigned to unvaccinated children in the objective function. Diminishing returns emerge as $\beta$ exceeds \$2.0, beyond which missed doses plateau, as the penalty term exceeds the average cost of a vaccine dose. Explicitly included in both UG and SB model objectives, this parameter also serves as an analytical metric, assessing total vaccination coverage, as compared to demand, for the system. While no true dollar amount can be placed on the long-term costs of missed vaccination, $\beta$ represents our attempt to capture and reduce those long-term implications.

\paragraph{Shared holding costs.}

Although holding costs arise at multiple points in the vaccine supply chain—including production, interim storage, and downstream distribution—these costs cannot be reliably disaggregated due to limited observability of producer-level inventories and country-level storage decisions from the coordinating entity level. We therefore do not model holding costs explicitly at each stage of the system. Instead, we incorporate a single aggregate holding cost term that serves as a reduced-form representation of storage and timing frictions throughout the supply chain. This captures the economic effect of delayed delivery without requiring detailed allocation of costs across producers or recipient countries. Holding cost in period $t$ is defined as $\sum_{v \in \mathcal{V}} h_v\bar{r}_{vt}I_{vt}$, the final term of \eqref{eq:UG_MP}.

\paragraph{Solve times.} Table~\ref{tab:solve_times} reports mean solve time and standard deviation across the 30 instances for each problem--objective combination.

\begin{table}[!tb]
\centering

\caption{Mean solve time (seconds) and standard deviation across 30 trials by objective/discount.}
\label{tab:solve_times}
\small{
\begin{tabular}{lrrrrrr}
\toprule
 & \multicolumn{2}{c}{\textbf{UG}} & \multicolumn{2}{c}{\textbf{SB}} & \multicolumn{2}{c}{\textbf{MP}} \\
\cmidrule(lr){2-3} \cmidrule(lr){4-5} \cmidrule(lr){6-7}
\textbf{Problem} & \textbf{Mean} & \textbf{SD} & \textbf{Mean} & \textbf{SD} & \textbf{Mean} & \textbf{SD} \\
\midrule
1: 1 segment                    & 53.27    & 28.30    & 21.50    & 10.42   & 40.03     & 25.67    \\
2: 2 segments                   & 485.41   & 227.42   & 132.70   & 49.83   & 355.05    & 239.59   \\
3: 2 segments, penta/hexa discount      & 516.32  & 167.63 & 180.77   & 61.10   & 437.66    & 98.75    \\
4: 2 segments, no penta/hexa    & 417.10   & 85.68    & 156.40   & 35.72   & 372.54   & 176.82  \\
5: 2 segments, HPV/PCV/Rota discount & 468.75 & 142.86 & 121.45 & 42.17 & 326.82 & 116.55    \\
6: 3 segments                   & 229.93   & 80.86    & 75.20    & 33.56   & 170.01    & 89.92    \\
\bottomrule
\end{tabular}}
\end{table}

\end{document}